\documentclass{article}
\usepackage{blindtext}
\usepackage{geometry}
\usepackage{graphicx} 
\usepackage{xcolor}
\usepackage{amsmath,amssymb}
\usepackage{cite}
\usepackage{url}
\usepackage{dsfont}
\usepackage{complexity}
\usepackage{amssymb}
\usepackage{dsfont}
\usepackage{physics} 
\usepackage{colortbl}
\usepackage{url}
\usepackage{hyperref}
\usepackage{subcaption}
\usepackage{authblk}

\usepackage[normalem]{ulem}
\newtheorem{definition}{Definition}[section]

\title{In Search of Quantum Advantage: Estimating the Number of Shots in Quantum Kernel Methods}
\date{}
\author[1]{Artur Miroszewski}
\author[2]{Marco Fellous Asiani}
\author[1]{Jakub Mielczarek}
\author[3]{Bertrand Le Saux}
\author[4,5]{Jakub Nalepa}
\affil[1]{Jagiellonian University, prof. S. \L{}ojasiewicza 11, 30-348 Cracow, Poland}
\affil[2]{University of Warsaw, Banacha 2c, 02-097 Warsaw, Poland}
\affil[3]{European Space Agency, Largo Galileo Galilei 1, 00044 Frascati, Italy}
\affil[4]{Silesian University of Technology, Akademicka 16, 44-100 Gliwice, Poland}
\affil[5]{KP Labs, Bojkowska 37J, 44-100 Gliwice, Poland}

\begin{document}

\maketitle

\begin{abstract}


Quantum Machine Learning (QML) has gathered significant attention through approaches like Quantum Kernel Machines. While these methods hold considerable promise, their quantum nature presents inherent challenges.
One major challenge is the limited resolution of estimated kernel values caused by the finite number of circuit runs performed on a quantum device.
In this study, we propose a comprehensive system of rules and heuristics for estimating the required number of circuit runs in quantum kernel methods. We introduce two critical effects that necessitate an increased measurement precision through additional circuit runs: the spread effect and the concentration effect. The effects are analyzed in the context of fidelity and projected quantum kernels. To address these phenomena, we develop an approach for estimating desired precision of kernel values, which, in turn, is translated into the number of circuit runs. Our methodology is validated through extensive numerical simulations, focusing on the problem of exponential value concentration. We stress that quantum kernel methods should not only be considered from the machine learning performance perspective, but also from the context of the resource consumption. The results provide insights into the possible benefits of quantum kernel methods, offering a guidance for their application in quantum machine learning tasks.
\end{abstract}

\section{Introduction}\label{sec:introduction}

The field of Quantum Machine Learning (QML\footnote{The table of abbreviations is available in Appendix \ref{app:abbreviations}.}) \cite{biamonte2017quantum, schuld2015introduction} has undergone an impressive development through the recent years, starting with general ideas, first proofs-of-concept, to actual small-scale implementations on quantum devices. 
Along the way, the community has identified both the possible opportunities related to specific algorithms as well as obstacles which need to be tackled to unleash the full potential of QML. One of the main research directions in QML are quantum kernel methods, specifically classical machine learning routines utilizing kernel matrices obtained with a Quantum Kernel Estimation (QKE) procedure \cite{havlivcek2019supervised, schuld2019quantum}.
Quantum kernel methods have been studied extensively from the machine learning task performance perspective for a plethora of use-cases, including finance \cite{miyabe2023quantum, grossi2022mixed}, Earth observation \cite{delilbasic2021quantum, miroszewski2023detecting}, drug discovery \cite{cao2018potential}, high-energy physics \cite{chan2021sissa}, and many more \cite{liu2021rigorous, vasques2023application, mengoni2019kernel}.

However, much less work so far has been performed to systematically assess the utility of quantum kernel methods from the perspective of the resources' consumption.
For the actual application of quantum devices for real-life problems, one does not necessarily need an algorithm which---in some theoretical limits---is proven to be the best, but rather a solution which is optimal from the perspective of a specified landscape of constraints and objectives.
Recently, an array of approaches to address this issue have emerged by defining a \textit{practical quantum advantage} of \textit{utility} of quantum algorithms \cite{herrmann2023quantum, hibat2023framework}. In short, the quantum algorithm is practically advantageous if it (\textit{i}) requires less computing time, (\textit{ii}) requires less energy, or (\textit{iii}) yields more accurate results.
In the case of quantum kernel methods, the points (\textit{i}) and (\textit{ii}) are inextricably linked to the number of quantum circuit runs (also called number of shots, $N$). Therefore, a practical set of rules for the number of shots estimation is essential, if one thinks seriously about performing QKE on real quantum devices. 


In this work, we propose a method to estimate sufficient number of shots, $\tilde{N}$, coming from, what we believe are, the two most important effects to consider when performing Quantum Kernel Estimation (QKE). The first effect---\emph{the spread effect}---pertains to the collective behavior of kernel values and can be likened to a signal-to-noise ratio within the kernel matrix. We aim to ensure that kernel values corresponding to different patterns in the data are well-separated, relative to the typical spread of kernel values in the matrix. The second effect---\emph{concentration avoidance}---addresses the tendency of quantum kernels to converge to a constant value as the size of the quantum device increases. Our goal is to distinguish kernel values from this convergent value.

The estimation of $\tilde{N}$, derived from the spread effect and concentration avoidance is obtained for each individual value in a kernel matrix. Knowing $\tilde{N}$ for each kernel value would provide complete knowledge of the kernel matrix, rendering the subsequent QKE procedure unnecessary. Therefore, in practice, it is essential to assume, \emph{a priori}, a fixed number of circuit runs in QKE for the entire dataset under consideration. We propose a set of heuristics to determine $N$ for the whole kernel matrix, $\bar{N}$, without full knowledge of the system, but by utilizing some statistical measures of the kernel matrix. This approach is particularly useful for planning real hardware experiments, especially if the system exhibits predictable statistical behavior.

As an example, we conduct a case study on the problem of exponential value concentration. Exponential value concentration imposes specific scaling laws on the collective behavior of kernel values as the size of the quantum device increases. This allows us to draw conclusions about the system from small, simulable instances and extend these conclusions to larger, classically intractable problem sizes.\\
Moreover, the proposed approach for $\tilde{N}$ estimation is analyzed both for the idealized noiseless computation and for noisy circuits with the possible error correction scheme.

\subsection{Contribution}

Understanding the (\textit{i}) and (\textit{ii}) aspects of quantum kernel methods is essential, due to the concentration issues \cite{thanasilp2022exponential} found in the popular quantum kernel families, including the fidelity \cite{havlivcek2019supervised} (FQ) and projected \cite{huang2021power} (PQ) quantum kernels. 
In general, due to the exponential kernel value concentration, as we increase the size of a quantum device, in order to distinguish kernel matrix values from one another, we are forced to increase a precision of QKE. This, in turn, exponentially escalates the number of quantum circuit runs we have to perform, and directly translates to the elevated runtime and energy consumption, making QKE infeasible for large quantum devices.
Therefore, the main research questions we would like to address in this work is as follows: 
\begin{itemize}
    \item \textbf{\textit{How to practically assess the number of required circuit runs for a quantum kernel estimation, that leads to correctly evaluated kernel values?}}
    \item \textbf{\textit{What size of a quantum device allows us to gain the practical quantum advantage over classical simulations, while still keeping the feasibility of the QKE?}}
\end{itemize}
There are results in the literature \cite{suzuki2022quantum} which indicate that not all kernel families suffer from the concentration problem. However, it is conjectured that avoiding concentration issues in QML leads to classical tractability \cite{cerezo2023does}. Hence, we investigate only the kernel families that are shown to suffer from the exponential value concentration.


\subsection{The Roadmap}

In Section~\ref{sec:theory}, we review quantum kernels, and introduce quantum feature maps. Then, we define and discuss a measurement process in two popular quantum kernel families: fidelity and projected quantum kernels and, finally define the exponential value concentration. In Section~\ref{sec:methods}, we focus on the resource estimation for QKE. The main part of this section is the discussion on how to infer the number of measurements $N$ needed to achieve distiguishability of different kernel matrix entries. In Section~\ref{sec:results}, we perform a numerical analysis of a ZZ-feature map~\cite{havlivcek2019supervised} based quantum kernels. We study both the \emph{expressibility} and \emph{entanglement} of the feature map for an example benchmark dataset targeting binary classification~\cite{sonar}--- we find the severity and consequences of the exponential concentration issue for both families of kernels. This allows us to estimate the number of measurements in both cases. We extend the analysis to both ideal (noiseless) and error-corrected circuits. We additionally present the results of a similar analysis for other widely-established machine learning datasets. In Section~\ref{sec:discussion}, we discuss the results and extend them by investigating the computational time and energy consumption for a classical simulation and QKE. We present the assumptions which are necessary for a meaningful analysis, but---at the same time---investigate the results which are \emph{overly optimistic} for quantum methods. Therefore, one should treat the results as the best case scenario, which is unlikely to occur in real-life problems.
Even with this na\"ive approach, we can rule out quantum device sizes which would not give rise to any benefit in using quantum computing. Section~\ref{sec:conclusions} concludes our findings.

\begin{figure}[ht!]
    \centering
    \includegraphics[width = 0.7\textwidth]{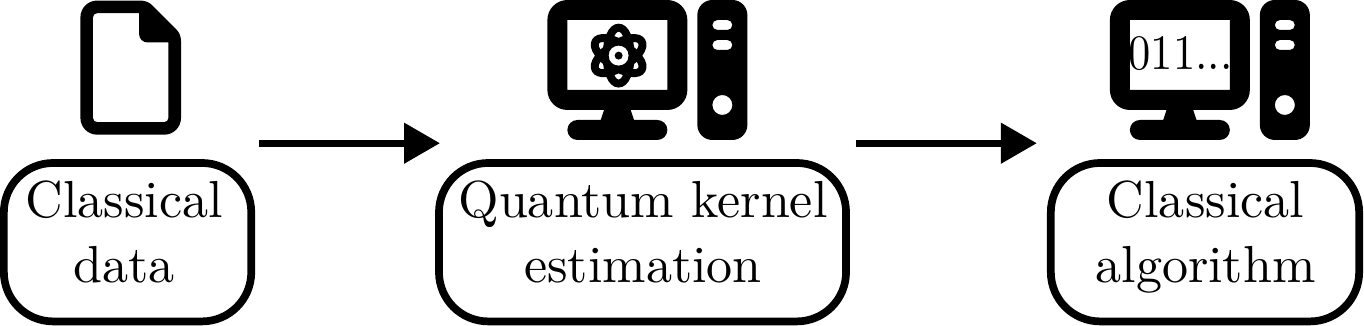}
    \caption{The general idea behind the quantum kernel machines. First, the classical or quantum data is encoded into the quantum circuits. The circuits are run, the kernel matrix entries or density matrices are estimated. For different quantum kernel families the subsequent classical post-processing might be necessary. Then, the results of a quantum kernel estimation serve as an input to classical routines.}
    \label{fig:hSVM}
\end{figure}

\section{Theoretical Background}\label{sec:theory}

\subsection{Quantum Kernel Methods}

Kernel methods are a class of algorithms used in machine learning for pattern analysis, both in supervised learning tasks like classification \cite{cortes1995support} and regression \cite{smola2004tutorial} and unsupervised learning, like dimensionality reduction or clustering \cite{scholkopf1998nonlinear}.
They map the input data into a higher-dimensional feature space where the data becomes linearly separable. The mapping is called a \textit{feature map}. Then, one calculates the similarity between pairs of embedded data points. This task can be facilitated by the \emph{kernel trick} \cite{berlinet2011reproducing}. Kernel methods are prized for their ability to handle non-linear relationships in data, and they have been proven extremely effective in an array of machine learning tasks \cite{larson2019review, hofmann2008kernel, apsemidis2020review}.

The idea of using kernels is crucial for quantum kernel methods~\cite{havlivcek2019supervised, schuld2019quantum}. In general, one encodes classical data into a quantum device, and performs quantum computation and measurements. The procedure is known as a Quantum Kernel Estimation (QKE). With the possible classical post-processing, the result is a kernel matrix which then can be used as an input for a classical optimization routines. The general idea behind quantum kernel methods is presented in Figure~\ref{fig:hSVM}. 
The motivation is two-fold.
On the one hand, it is believed that some kernel functions, especially the ones which are induced by entangling quantum feature maps, can be obtained with fewer resources on a quantum computer \cite{preskill2018quantum, havlivcek2019supervised, goldberg2017complexity}. That includes algorithms' runtime or energy consumption \cite{schuld2019quantum}.
On the other hand, the possibility of embedding data points in exponentially growing (with the number of qubits used) feature spaces, gives hope for building machine learning models which can represent a wide range of functions or patterns in the data \cite{havlivcek2019supervised, huang2021power}.

\subsection{Quantum Kernels}

\subsubsection{Data Embedding}
In order to process the classical data on a quantum computer, one has to encode the data into the quantum device.
A general way to embed the classical datum $x$ into the quantum state is to apply a parameterized unitary transformation $U: \mathcal{X} \mapsto \mathbb{U}_x \subset \mathcal{U}(d)$, where $\mathcal{U}(2^n)$ is the total space of unitaries of dimension $2^n$, on an initial $n$-qubit quantum state ($|0\rangle^{\otimes n} \equiv |0\rangle$):
\begin{equation}\label{eq:embedded_state}
    \rho(x) = U(x) \rho_0 U^{\dagger}(x),\ \rho_0 = | 0 \rangle \langle 0 |,
\end{equation}
where $U^{\dagger}$ represents a Hermitian conjugate of the operator $U$.

\noindent In general, it is typical in quantum kernel methods to parameterize the embedding unitary transformation with additional parameters $U(x,\Theta)$. This is done to adjust the feature map in such a way that the resulting kernel function is compatible with the learning task \cite{hubregtsen2022training}. In this work, we assume that we have access to the already well-trained quantum feature map, therefore we can limit ourselves to the feature maps parameterized only with data $U(x) \in \mathbb{U}_x$. The two characteristics that influence the collective behavior of kernel entries (see, kernel value concentration in Section \ref{sec:exponential_value_conc}) and depend both on the data and quantum embedding are \emph{expressibility} and \emph{entanglement}, introduced in Appendix \ref{app:characteristics}. Generally speaking, \emph{expressibility} is the ability of the model to represent a wide range patterns in the data, while \emph{entanglement} characterizes how much data is represented in the correlations between subsystems of the quantum device.

\subsubsection{Fidelity Quantum Kernel}

With the classical data embedded into quantum states, one can use a quantum device to estimate a measure of similarity between states $\rho(x)$ and $\rho(y)$. 
This measure of similarity is dubbed a \emph{quantum kernel}. 
Two types of quantum kernels are extensively studied in the literature: the fidelity~\cite{havlivcek2019supervised, schuld2019quantum} and projected quantum ones~\cite{huang2021power}. 
While discussing the two kernel families below, we fix our attention on the non-SWAP test implementations of quantum circuits.
The original kernel function proposed for quantum kernel support vector machines is the fidelity kernel \cite{havlivcek2019supervised, schuld2019quantum}. 
It is defined as the squared overlap between the states $\rho(x)$ and $\rho(y)$:
\begin{equation}\label{eq:fidelity_kernel}
\kappa^{FQ}(x,y) = \Tr[\rho(x) \rho(y)].
\end{equation}

\noindent From the practical perspective, the fidelity kernel's value can be estimated by projecting the encoded-decoded data points $U^{\dagger}(y) U(x) \rho_0 U^{\dagger}(x) U(y)$ on the vacuum state $\rho_0 = (|0\rangle \langle 0|)^{\otimes n}$:
\begin{equation}\label{eq:fidelity_enc_dec}
    \kappa^{FQ}(x,y) = \Tr [  U(y) \rho_0 U^{\dagger}(y) U(x) \rho_0 U^{\dagger}(x)] = \Tr [ \rho_0 U^{\dagger}(y) U(x) \rho_0 U^{\dagger}(x) U(y)].
\end{equation}

\noindent The result of a single circuit run can be either a ``success", when the encoded state, during the measurement, collapsed to the vacuum state or, a ``failure", when the encoded state collapsed to any other state. 

\begin{figure}[ht]
\centering
    \includegraphics[width=0.8\textwidth]{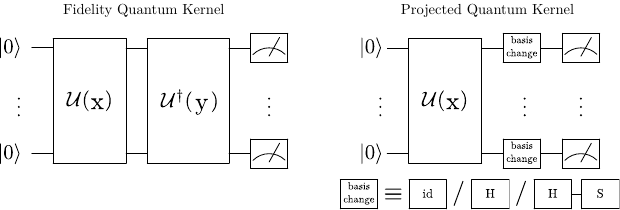}
    \caption{Schematic quantum circuits used for quantum fidelity and projected kernel estimation. \textit{Left:} Fidelity quantum kernel circuit for the concatenation method of the kernel value estimation. Two data points $x,y$ are encoded-decoded with the embedding map unitary transformation $U(\cdot)$. The outcome of the measurement is used for gathering the final vacuum/non-vacuum state statistics. \textit{Right:} Projected quantum kernel circuit for the one-qubit quantum state tomography. First, the data point $x$ is encoded with the feature map unitary transformation $U(x)$, producing full $n$-qubit embedded state. Then, for each of the qubit registers,the one-qubit reduced density matrix is measured in different bases. After performing quantum tomography of the reduced states, the estimated density matrices are processed further classically in order to obtain projected kernel values.}
    \label{fig:kernel_implementation}
\end{figure}

The practical implementation of the fidelity quantum kernel estimation in quantum computers is usually performed by the oschmidt Echo test (also called the circuit concatenation method)(Figure~\ref{fig:kernel_implementation}), but also other approaches exist. Firstly, the data point $x$ is encoded with embedding map unitary transformation $U(x)$. Then another point $y$ in encoded with the inverse embedding map unitary transformation $U^{\dagger}(y)$ and the state is measured. 
The QKE of fidelity kernel is effectively the frequency of obtaining a vacuum state $\rho_0$ in a series of measurements, see Section \ref{sec:bernoulli}.

\subsubsection{Projected Quantum Kernel}

The second type of a quantum kernel, we study, is the projected quantum kernel \cite{huang2021power}.
Generally, a projected quantum kernel is a semi positive-definite function of the reduced density matrices of the quantum system.
From the theoretical perspective, the reduced density matrices are obtained by tracing out some degrees of freedom from the full density matrix. 
The one-qubit reduced density matrix will be denoted as $\rho_k$, where $k$ is the index of the qubit register which was not traced out, $\rho_k(x) = \Tr_{j\neq k}[\rho(x)]$.
In practice, to obtain the reduced density matrix one has to perform repeated measurements of the embedded state $\rho$ in different bases and use the results from the qubit registers in a process called the \emph{quantum state tomography} \cite{d2003quantum}. The enhanced tomography schemes, like classical shadows \cite{huang2021power, gosset2018compressed, aaronson2018shadow} can be applied here.
The obtained reduced density matrices are then used in classical post-processing to obtain a kernel value.
The function used in classical post-processing can have a wide variety of forms, both linear and non-linear, with one-qubit and multiple-qubit functions being notable examples \cite{huang2021power}.
In this study, we focus on the Gaussian kernel function using one-qubit reduced density matrices, as it is popular in the literature and, contrary to the fidelity quantum kernel case, it needs introducing additional approximations in the presented methods (Section \ref{sec:methods}). This way, the reader is fully equipped to apply proposed methods to other projected quantum kernel functions.
Following the original definition of a Gaussian kernel function \cite{huang2021power}, we define
\begin{equation}\label{eq:projected_kernel}
    \kappa^{PQ}(x,y) = \exp(- \gamma \sum_k ||\rho_k(x) - \rho_k(y)||^2_2).
\end{equation}
In this kernel type, we have to first perform a quantum one-qubit state tomography to obtain $\rho_k(x)$ and $\rho_k(y)$ matrices and then perform classical post-processing.
The one-qubit density matrix consists of two diagonal entries $\rho_{00}$ and $\rho_{11}$ which are real numbers which add up to one: $\rho_{00} + \rho_{11} = 1$ and two complex off-diagonal entries $\rho_{01}$ and $\rho_{10}$, which are related by a complex conjugation $\rho_{01} = \rho_{10}^*$. Therefore, we ultimately need to find three independent real numbers to gain the full information about the one-qubit density matrix. 

Let us choose the three independent variables to be: $\rho_{00}, \Re(\rho_{01}), \Im(\rho_{01})$.
To estimate these variables, one has to perform a state measurement in three different bases. For example: the standard computational Z-basis $\{| 0 \rangle, |1\rangle\}$, Hadamard rotated X-basis $\{|+\rangle = \hat{H}| 0 \rangle, |-\rangle = \hat{H}| 1 \rangle \}$, and Hadamard-Phase rotated Y-basis $ \{|+i\rangle = \hat{S}\hat{H}| 0 \rangle, |-i\rangle = \hat{S}\hat{H}| 1 \rangle \}$, see Figure \ref{fig:kernel_implementation}.
To stay consistent with the approach presented in this paper, we perform the projection on the vacuum state. For a one-qubit reduced state $\rho_k(x) = \Tr_{j \neq k}[U(x)\rho_0 U^{\dagger}(x)]$, we have
\begin{equation}\label{eq:projected_rhos}
    \begin{split}
        \rho^D_k & = \rho_{00,k} = \Tr[|0\rangle \langle 0 | \rho_k(x)] = \mathbb{E}[\hat{\mathcal{M}}^D], \\
        \rho^R_k & = \Re(\rho_{01,k}) = \Tr[|+\rangle \langle + | \rho_k(x)] - \frac{1}{2} = \Tr[|0\rangle \langle 0 | \hat{H} \rho_k(x) \hat{H}] - \frac{1}{2} = \mathbb{E}[\hat{\mathcal{M}}^R] - \frac{1}{2}, \\
        \rho^I_k & = \Im(\rho_{01,k}) = \frac{1}{2} - \Tr[|+i\rangle \langle +i| \rho_k(x)] = \frac{1}{2} - \Tr[|0\rangle \langle 0|\hat{H}\hat{S}^{\dagger} \rho_k(x) \hat{S} \hat{H}] = \frac{1}{2}-\mathbb{E}[\hat{\mathcal{M}}^I].
    \end{split}
\end{equation}
\subsection{Exponential Value Concentration}\label{sec:exponential_value_conc}
Here, we follow the definition of the exponential concentration for quantum kernels given in \cite{thanasilp2022exponential}.

\begin{definition}[Exponential kernel value concentration]
Consider independent quantum kernel values $\kappa(x,x')$ that depend on the data $x, x'$ and are measured from a quantum computer as the expectation value of some observables.
$\kappa(x,x')$ is said to be deterministically exponentially concentrated in the number of qubits $n$ towards a certain fixed value $\mu$ if
\begin{equation}\label{eq:deter_exp_conc}
    |\kappa(x,x') - \mu| \leq \beta \in O(1/b^n),
\end{equation}
for some $b>1$ and all $x,x'$.
Analogously, $\kappa(x,x')$ is probabilistically exponentially concentrated if
\begin{equation}\label{eq:prob_exp_conc}
    Pr_{x,x'}\left[ |\kappa(x,x')-\mu| \geq \delta \right] \leq \frac{\beta^2}{\delta^2}, \beta \in O(1/b^n),
\end{equation}
for $b>1$.
That is, the probability that $\kappa(x,x')$ deviates from $\mu$ by a small amount $\delta$ is exponentially small for all $x,x'$.

\end{definition}

\subsection{Measurement of Quantum Kernels as a Bernoulli Process}\label{sec:bernoulli}
Observe that both in the case of the fidelity quantum kernel and projected quantum kernel, a single measurement gives as an outcome a single of two possible values. 
Although for the fidelity quantum kernel, the measured state can be any eigenstate from the $2^n$ dimensional Hilbert space, in the kernel estimation, we interpret it either as an initial vacuum state (``success") or not (``failure").
Employing the Born rule, the probability of ``success" is $\Tr \left[ \rho(x) \rho(y) \right]$ which is the true kernel value $\kappa^{FQ}(x,y)$. By performing multiple circuit runs, we obtain ``success" vs. ``failure" statistics which can be used for the quantum kernel (maximum likelihood) estimation:
\begin{equation}\label{eq:fidelity_binomial}
    \hat{\kappa}^{FQ}(x,y) = \frac{1}{N} \sum_{i=1}^N \hat{X}_i,
\end{equation}
where $\hat{X}_i \in \{0 - ``{\rm failure}", 1 - ``{\rm success}"\}$ is the outcome of the $i^{th}$ circuit run and $N$ is the number of circuit runs of a quantum device. In the asymptotic limit of the number of shots, the measured kernel value tends to the true value, $\hat{\kappa}^{FQ}(x,y)\xrightarrow[]{N\rightarrow\infty} \kappa^{FQ}(x,y)$. When we treat $\hat{X}_i$ as a random variable, the fidelity quantum kernel estimation is equivalent to the proportion estimation in binomial distribution $Bin(\hat{k};N, p=\kappa^{FQ}(x,y))$, where $\hat{k}$ is the number of ``successes" $\hat{k} = \sum_{i=1}^N \hat{X}_i$ drawn in $N$ trials with a single trial success probability $p=\kappa^{FQ}(x,y)$. 
For the fidelity quantum kernel, the measured values will concentrate to $\mu^{FQ}=0$. This results in a poor success rate of a measurement --- a small chance of obtaining even a single ``success" in all $N$ trials.
The understanding of the effect of the exponential value concentration can be simply explained with random states on the quantum Hilbert space. If the embedding map for the points were composed of Haar random unitaries, then the points would be randomly distributed in the exponentially large number ($2^n$) of orthogonal directions in the Hilbert space. Hence, in the case of the fidelity kernel, the chance that two data points have the non-vanishing components in the same direction and their scalar product is non-zero is exponentially small. The expectation value of such kernel with respect to the random unitaries, as a function of a number of qubits $n$ is $\langle \kappa_{Haar} \rangle = 2^{-n}$, while the standard deviation of such random kernels is $\sigma(\kappa_{Haar}) = \langle \kappa_{Haar}^2 \rangle - \langle \kappa_{Haar} \rangle^2 = 2^{-n}\sqrt{(1-2^{-n})/(1+2^{-n})} \xrightarrow{n\rightarrow \infty } 2^{-n}$. 
In order to obtain a full fidelity kernel matrix for a given problem consisting of $m$ data points, we need to estimate $\frac{m(m-1)}{2}$ independent kernel values. That directly translates into 
$N_{total}^{FQ} = \frac{N}{2} m(m-1)$
circuit runs.

Contrary to the fidelity quantum kernels, for projected quantum kernels one has to resort to the quantum tomography of states and then to classical post-processing.
Nevertheless, for the quantum tomography of independent density matrix elements, the measurement interpretation as a Bernoulli process can be extended to a fidelity kernel measurement for a reduced-density matrix (see Eq. \ref{eq:projected_rhos}).
For a single one-qubit reduced density matrix, we will have to perform $3 \cdot N$ measurements.
Observe that for a single projected kernel value (Eq.~\ref{eq:projected_kernel}), one has to perform a quantum state tomography for all $n$ one-qubit density matrices $\{\rho_k(x)\}_{i=1}^{n}$. 
Fortunately, all one-qubit density matrices can be estimated in parallel. One takes the $n$-qubit quantum computer, perform quantum data embedding $\rho(x) = U(x)\rho_0 U^{\dagger}(x)$, and measure all qubits in the chosen basis. 
The $k^{th}$ bit in the resulting bitstring will be a single trial of a binomial random variable for estimation of $\rho_k(x)$ density matrix.
For a highly entangling feature maps, the reduced-density matrices $\rho_k(x)$'s will tend toward the maximally mixed state. For a one-qubit reduced density matrix, the independent variables will concentrate to $\rho_k^D \rightarrow 1/2, \rho_k^R \rightarrow 0, \rho_k^I \rightarrow 0$. Of note, the actual measured expressions are described with the random variables $\hat{\mathcal{M}}_k^D = \Tr[|0\rangle \langle 0 | \rho_k(x)], \hat{\mathcal{M}}_k^R =  \Tr[|+\rangle \langle + | \rho_k(x)], \hat{\mathcal{M}}_k^I = \Tr[|+i\rangle \langle +i| \rho_k(x)]$ with the binomial probability distribution $Bin(\hat{\mathcal{M}}_k^X,N,p=\mathbb{E}[\hat{\mathcal{M}}_k^X])$, $X=D,R,I$, respectively. Then those values are post-processed to obtain $\hat{\rho}_k^D, \hat{\rho}_k^R, \hat{\rho}_k^I$, see Eq. \ref{eq:projected_rhos}. Taking this into account, from the perspective of the measured expressions $\hat{\mathcal{M}}_k^X$, the concentration value for each of the measured value is $\mu^{PQ} = 1/2$. This, in turn, will not cause problems with the success rate of measurements but will impede the ability to distinguish whether the measured values are greater or less than $\mu^{PQ}$.
To find the kernel matrix, one performs quantum state tomography for all $m$ data points. Thus, for obtaining the full kernel matrix one has to perform 
$N_{total}^{PQ} = 3mN$,
runs of a quantum circuit.

\begin{figure}[ht]
\centering
    \includegraphics[width=0.7\textwidth]{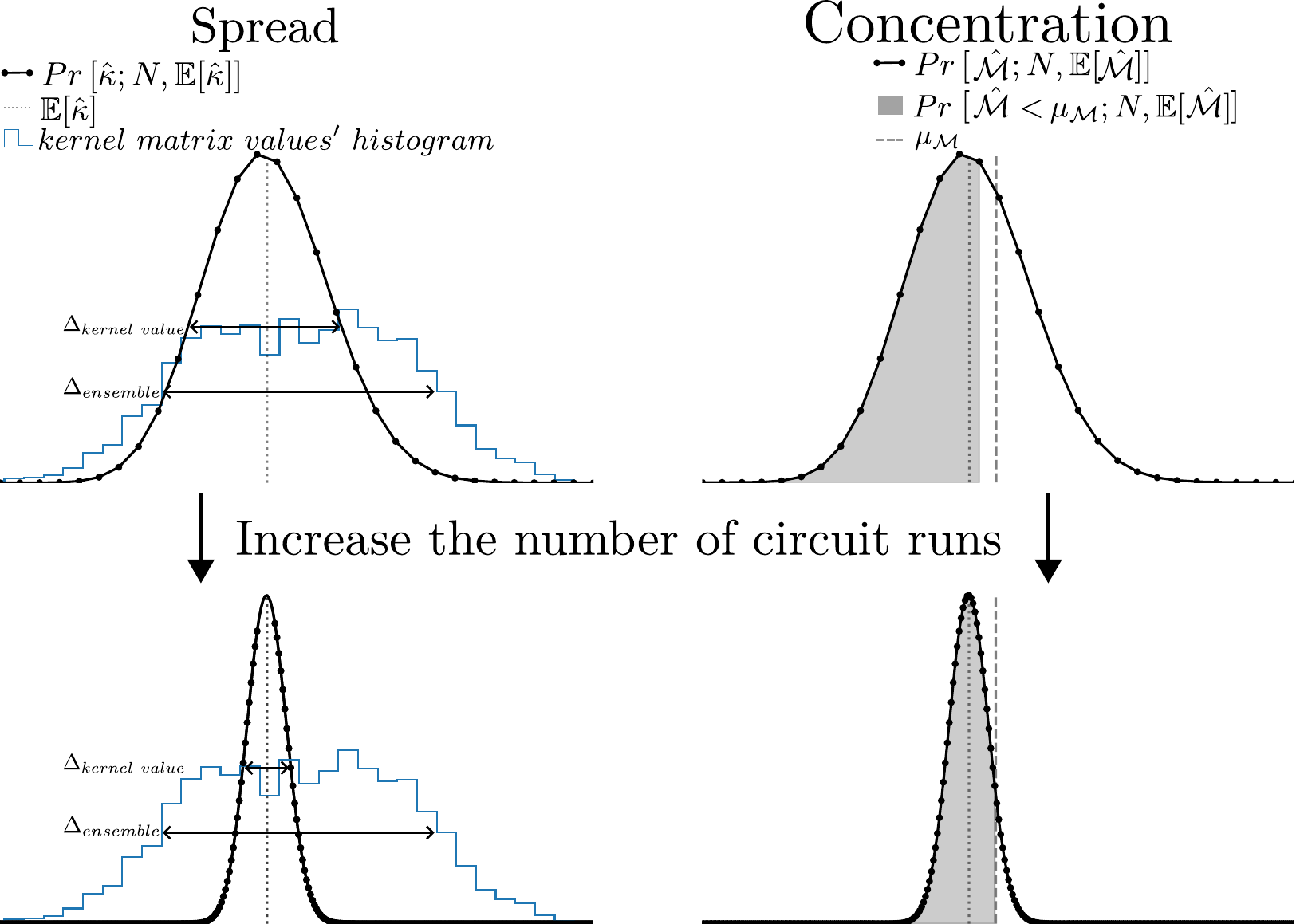}
    \caption{Schematic representation of the two effects taken into account during the number of shots, $N$, estimation.  \textit{Left:} The comparison between the distribution of single quantum kernel measurement with the histogram of the independent kernel values in the kernel matrix. The uncertainty of the kernel value estimation $\Delta_{kernel\ value}$ is similar to the spread of the independent kernel values in the kernel matrix $\Delta_{ensemble}$. Increasing $N$, reduces the $\Delta_{kernel\ value}$ allowing for distinguishing the single kernel values in the ensemble. \textit{Right:} The uncertainty related to the measurement is big enough to face challenges with distuingishing whether measured value $\hat{\mathcal{M}}$ is less or greater than the concentration value $\mu_{\mathcal{M}}$. Increasing $N$ allows to boost the probability of measuring the $\hat{\mathcal{M}}$ value being smaller than the concentration value $\mu_{\mathcal{M}}$.\\
    The plot is created for simulated data, the distributions are rescaled to improve the clarity of the reception of the figure.}
    \label{fig:N_effects}
\end{figure}

\section{Methods: Quantum Resource Estimation}\label{sec:methods}
As stated in the Section \ref{sec:introduction}, the main objective of this paper is to propose a system of rules for the evaluation of resources needed for the successful quantum kernel estimation procedure.
In the view of the exponential kernel concentration, being able to gauge the number of quantum circuit runs $\tilde{N}$ needed for a successful algorithm execution is of paramount importance and becomes the first step for assessing the use of resources, like runtime and energy expenditure. To estimate the sufficient number of shots $\tilde{N}$, we propose to investigate the effects connected to the spread of kernel values in the dataset and the concentration avoidance. 
The estimated number of shots $\tilde{N}_{spread}$ (formally introduced in the next section) is connected with the ability of distinguishing the single kernel value obtained in QKE procedure from other kernel values in the kernel matrix.
The estimated number of shots $\tilde{N}_{CA}$ (formally introduced in Section \ref{sec:CA}) stem from the observation for the fidelity quantum kernels that with the strong concentration of kernel values, the success rate of the measurement process vanishes, necessitating additional circuit runs.
This observation is then extended to projected quantum kernels.
After obtaining the estimates $\tilde{N}_{spread}$ and $\tilde{N}_{CA}$, one can decide on the number of circuit runs $\tilde{N}$ by, for example, taking the greater value to satisfy conditions for both effects:
\begin{equation}
    \tilde{N} = max(\tilde{N}_{spread},\tilde{N}_{CA}).
\end{equation}

\subsection{Spread of Kernel Values}
Being able to distinguish two kernel matrix values from one another is of crucial importance in any kernel method.
Therefore, one has to perform the comparison of two values: the uncertainty in estimation of a single kernel entry, and the spread of the distribution of kernel values in the kernel matrix. 
To distuinguish two kernel entries from one another, the uncertainty connected with the determination of a single kernel entry has to be comparable or, preferably, smaller than the typical difference between kernel values in the ensemble,

\begin{equation}\label{eq:precision_ratio}
   \varepsilon = \frac{\Delta_{kernel\ value}}{\Delta_{ensemble}}  \lesssim 1 .
\end{equation}

\noindent We denote the ensemble of kernel values as $\mathbb{K} = \{\kappa(x,y)| \forall x,y \in \mathcal{X} \}$.

Following Appendix III in \cite{thanasilp2022exponential}, one can estimate the sufficient number of shots connected to the effect of the spread in kernel values $\tilde{N}_{spread}$. It can be done by identifying $\Delta_{kernel\ value} = |\hat{\kappa} - \mathbb{E}[\hat{\kappa}]|$ and introducing an upper bound $P_{spread}$ on the probability that the estimated kernel value $\hat{\kappa}$ deviates strongly from the true $\kappa$ value, 
\begin{equation}
Pr \left[ |\hat{\kappa}-\mathbb{E}[\hat{\kappa}]| \geq \varepsilon\Delta_{ensemble}\right].\end{equation}
Using the Chebyshev's inequality and Eq.~\ref{eq:precision_ratio}, one finds that (see Appendix \ref{App:N_estimates})
\begin{equation}\label{eq:Spread}
    N \geq \frac{\mathcal{G}^{FQ/PQ}}{(1-P_{spread}) \varepsilon^2 \Delta_{ensemble}^2},
\end{equation}
where $P_{spread}$ is a probability of finding $\hat{\kappa}$ at least $\varepsilon\Delta_{ensemble}$-close to the true value $\kappa = \mathbb{E}[\hat{\kappa}]$, $\mathcal{G}^{FQ} = \mathbb{E}[\hat{\kappa}^{FQ}](1-\mathbb{E}[\hat{\kappa}^{FQ}])$ and $\mathcal{G}^{PQ} = \gamma^2 (\kappa^{PQ})^2 n \sum_k V_k$, with $V_k$ defined in Appendix \ref{App:N_estimates}.
Therefore, we define the sufficient number of shots for the estimation a single kernel entry to be the smallest value that satisfies the inequality (\ref{eq:Spread}),
    $N^{FQ/PQ}_{spread} = \mathcal{G}^{FQ/PQ}/(1-P_{spread}) \varepsilon^2 \Delta_{ensemble}^2.$

In the original derivation of the scaling law governing the minimal number of shots $N$ \cite{thanasilp2022exponential}, the spread of kernel values in the ensemble $\Delta_{ensemble}$ was measured with the use of standard deviation. 
Although this choice does not influence the qualitative result of the scaling of $N$, we argue that inter-quartile range (IQR) is a more sensible choice.
Firstly, in the case of the exponential value concentration for some kernel families, we deal with highly-skewed distributions for which the variance based measures lose their standard interpretation, stemming from the analysis of the normal distribution. 
Secondly, since IQR is based on the middle $50\%$ of the data, it is less influenced by the extreme values in comparison to the range or standard deviation.
Therefore, we propose to describe the spread of kernel values in the ensemble using IQR:
\begin{equation}
\Delta_{ensemble} = {\rm IQR}(\mathbb{K}).
\end{equation}

\subsection{Concentration Avoidance}\label{sec:CA}

The following effect is connected to the measurements performed during the single kernel value estimation procedure, rather than to the collective relations between kernel values in the ensemble.
Therefore, we focus on the measurement outcomes rather than on the actual kernel values.
For different families of kernels, the manifestation of the kernel concentration can lead to different obstacles in obtaining a reliable kernel value.

In the case of the fidelity quantum kernel, the decreasing value of the kernel value significantly reduces the success rate of the measurement.
It might happen that with the insufficient number of circuit runs $N$, all performed trials in the Bernoulli process will end up as being a ``failure".
Hence, we would like to increase the probability of at least one ``success'' in the measurement process, by increasing $N$. 

In the case of projected quantum kernel, the measurement values will, on average, tend to the equal number of ``failures" and ``successes" in the Bernoulli process. The success rate in kernel estimation will be high, but one will have a problem in distinguishing whether the true measured value $\mathbb{E}[\hat{\mathcal{M}}]$ is greater or less than the concentrated value. For the true measured value different from the concentration value, we wish to increase the precision of the kernel estimation procedure. This would result in ability to accurately tell whether the measured value is greater or less than the concentration value, see Fig. \ref{fig:N_effects}.


Gathering both of the above remarks into one, we propose a condition on the circuit runs number $N$ that for the true measured value $\mathbb{E}[\hat{\mathcal{M}}]$, with at least probability $P_{CA}$, the measured number of ``successes" $\hat{\mathcal{M}} = \frac{\hat{k}}{N}$ will be on the correct side of the concentrated value $\mu_{\mathcal{M}}$. By that, we mean, if the true measured value is less (greater) than the concentration value, we wish to increase the chance for the QKE to result in a smaller (greater) measured value. More formally,

\begin{equation}
    \begin{cases}\label{eq:sr}
        Bin(\hat{k} < N \mu_{\mathcal{M}},N,\mathbb{E}[\hat{\mathcal{M}}] ) \geq P_{CA}, & \text{for } \mathbb{E}[\hat{\mathcal{M}}] < \mu_{\mathcal{M}},\\
        Bin(\hat{k} > N \mu_{\mathcal{M}},N,\mathbb{E}[\hat{\mathcal{M}}] ) \geq P_{CA}, & \text{for } \mathbb{E}[\hat{\mathcal{M}}] > \mu_{\mathcal{M}}.
    \end{cases}
\end{equation}

\noindent In the case of the fidelity quantum kernel, the above bounds simplify to
\begin{equation}\label{eq:N_CA_FQ}
    N \geq log_{1-\mathbb{E}[\hat{\mathcal{M}}]}(1-P_{CA}),
\end{equation}
where above $1-\mathbb{E}[\hat{\mathcal{M}}]$ is the base of the logarithm. Similarly to the spread effect case we define the sufficient number of shots for concentration avoidance effect to be the smallest number satisfying the bound (\ref{eq:sr}). For the fidelity kernel case we have $N^{FQ}_{CA} = log_{1-\mathbb{E}[\hat{\mathcal{M}}]}(1-P_{CA})$.
Due to the straightforward relation between measured value and the form of the fidelity kernel value $\hat{\mathcal{M}} = \kappa^{FQ}$ the condition above (\ref{eq:sr}) equivalent to the condition
\begin{equation}
    Pr \left[ \hat{\kappa}^{FQ}(x,y) > 0 \right] \geq P_{CA},
\end{equation}
for the kernel value $\kappa^{FQ}(x,y)$ measured for the $x,y$ data points.

In the case of the projected quantum kernel, the formula (\ref{eq:sr}) cannot be simplified easily. On the other hand, here the approximation of the Bernoulli process with a normal probability distribution is justified. With the concentration value $\mu_{\mathcal{M}}$ not being on the edge of the measured value $\hat{\mathcal{M}}=\frac{\hat{k}}{N}$ range and for $N \gg 1$, one obtains
\begin{equation}\label{eq:PQ_AC}
    N \geq \frac{z^2 \mathbb{E}[\hat{\mathcal{M}}](1-\mathbb{E}[\hat{\mathcal{M}}])}{(\mathbb{E}[\hat{\mathcal{M}}]-\mu_{\mathcal{M}})^2},
\end{equation}
where $z = \Phi_N^{-1}(P_{CA})$ is the z-score and $\Phi_N^{-1}(P_{CA})$ is the inverse cumulative distribution function of the normal distribution evaluated at the probability $P_{AC}$. Again, we define the sufficient number of shots for this case to be $\tilde{N}_{CA}^{PQ} = z^2 \mathbb{E}[\hat{\mathcal{M}}](1-\mathbb{E}[\hat{\mathcal{M}}])(\mathbb{E}[\hat{\mathcal{M}}]-\mu_{\mathcal{M}})^2$.


\subsection{Noisy Computation}
All the above considerations investigated the uncertainty of the quantum kernel estimation procedure connected to the randomness of the measurement process, hence assumed noiseless circuits. 
The quantum computation inseparably includes noise from the hardware deficits.
Here, we propose a simple approach to analyze the consequences of noise in the previous estimates. This allows us also to expand our considerations by the error-correction procedures.
Assume that the noiseless circuit run should output a state of the system $\rho$. Due to the noise, the final state of the system might differ from the expected noiseless state. 
We introduce a probability of the error occurring during the execution of the circuit as $p$ and describe the final state of noisy computation as
\begin{equation}
    \rho_f = (1-p)\rho + p \Tilde{\rho},
\end{equation}
where $\Tilde{\rho}$ is an unspecified state obtained due to the noise occurrence.
During the quantum kernel estimation, we will then deal with the measurements of the type $\Tr[U| 0 \rangle \langle 0 | U^{\dagger} \rho_f]$, where $U$ is a possible measurement basis change.
For each circuit run, the resultant $\rho_f$ will be random, and in general, different. 
The kernels computed for $\rho_f$ instead of $\rho$ will be denoted with a subscript $f$, e.g., $\kappa_f^{FQ} = \Tr[|0\rangle \langle0|\rho_f]$.
First, we begin with the spread of kernel values assumption, which can be decomposed into two parts
\begin{align}
|\hat{\kappa}_f - \kappa| = |(\hat{\kappa}_f - \kappa_f)+(\kappa_f - \kappa)| \leq |\hat{\kappa}_f - \kappa_f|+|\kappa_f - \kappa| \leq \varepsilon\Delta_{ensemble}.\label{eq:k_difference_noisy}
\end{align}
We impose the following conditions to be true:
\begin{align}
    &|\hat{\kappa}_f - \kappa_f| \leq \frac{\varepsilon\Delta_{ensemble}}{2},\label{eq:SR_noise} \\ 
    &|\kappa_f - \kappa| \leq \frac{\varepsilon\Delta_{ensemble}}{2},\label{eq:SR_error}
\end{align}
where the first inequality takes over the kernel spread condition for noiseless circuits, while the second condition restricts the amount of permitted error rate $p$, see Appendix \ref{App:noisy_circuits_new}.

Starting with the latter, we find that
\begin{equation}
    p \leq \frac{\varepsilon \Delta_{ensemble}}{2\Delta_f^{FQ/PQ}},
\end{equation}
where $\Delta_f^{FQ} = |\Tr[|0\rangle \langle 0|\Tilde{\rho}-\mathbb{E}[\hat{\kappa}^{FQ}]]|$, $\Delta_f^{PQ} = 2\gamma\kappa |\sum_k(||\Delta \rho_k||^2_2)-Re[\langle \Delta\rho_k, \Delta\Tilde{\rho}_k \rangle_F] |$ and the result for projected quantum kernels is a first order approximation in $p$. With the assumption of the depolarizing noise channel one obtains significant simplifications $\Delta_f^{FQ} = 2^{-n}-{\kappa}^{FQ}$ and $\Delta_f^{PQ}=-2\kappa^{PQ} \ln \kappa^{PQ}$. This condition can be useful for error budget estimation in error correction schemes.

The condition (\ref{eq:SR_noise}) translates to
\begin{align}
    N^f_{spread} = \frac{4\mathcal{G}_f^{FQ/PQ}}{(1-P_{spread}) \varepsilon^2 \Delta_{ensemble}^2},\label{eq:spread_noisy}
\end{align}
where $\mathcal{G}_f^{FQ} = 1$ and $\mathcal{G}_f^{PQ} = \gamma^2 (\kappa_f^{PQ})^2 n \sum_k V^f_k$, with $V^f_k$ defined in Appendix \ref{App:noisy_circuits_new}. The main difference from the noiseless case is that we cannot use the variance expression for the binomial distribution and the bound is less stringent.
The factor of $4$ in the numerator comes from dividing the condition (\ref{eq:k_difference_noisy}) into (\ref{eq:SR_noise}) and (\ref{eq:SR_error}) while halving the allowed spread of kernel values.

For the considerations concerning the concentration avoidance effect we should emphasize that with the noisy circuits each circuit run is a random process with $\rho_f$ treated as a random variable.
Therefore, the measurement process will not be described with the binomial distribution, which is a composition of separate Bernoulli trials with respect to the same probability. Nevertheless, one can easily include noisy circuits in the analysis by extending the considerations to the collection of subsequent Bernoulli trials with probability $\mathbb{E}[\hat{\mathcal{M}}^f]$ treated as a random variable. For the considerations connected to the concentration avoidance effect, one finds that that the form of the conditions obtained for the noiseless case carry over to the noisy case. The single difference is that instead of $\mathbb{E}[\hat{\mathcal{M}}]$ one has to consider $\mathbb{E}[\hat{\mathcal{M}^f}]$. The details of the analysis can be found in Appendix \ref{App:noisy_circuits_new}.
As an example, the condition of the noisy fidelity quantum kernel becomes,
\begin{equation}\label{eq:N_CA_FQ_F}
    \tilde{N}_{CA}^{FQ} = log_{1-\mathbb{E}[\hat{\kappa}_f]}(1-P_{CA}).
\end{equation}

\subsection{Determining $N$ for the Whole Dataset}
Determining the number of circuit runs, $N$, for the problems in which classical emulation of the quantum kernels is feasible, does not serve, other than benchmarking, a useful purpose. Therefore, obtaining $\tilde{N}$ for every single independent kernel value in the kernel matrix is meaningless in practical applications.
On the other hand, if one knows a statistical behavior of kernel values in the kernel matrix, then one is in a position to assign specific $N$ to the whole dataset for a given setup, we denote this value as $\bar{N}$. The chosen $\bar{N}$ should, on average, provide sufficient kernel value estimation precision in order not to obstruct the target kernel task. 
Here again we decide on the number of circuit runs $\bar{N}$ for the whole kernel matrix by
taking the greater value to satisfy conditions for spread and concentration avoidance effects effects:
\begin{equation}
    \bar{N} = max(\bar{N}_{spread},\bar{N}_{CA}).
\end{equation}
In the paragraphs below, we will focus our attention on the classification task.

For the spread effect, we propose to determine $\bar{N}_{spread}$ based on the statistical measures connected to percentiles. The measures should also take into account the machine learning task performed by kernels. For example, for a balanced binary classification problem, $\frac{m(m-1)}{2}$ independent kernel values will consist of $\left(\frac{m}{2}\right)^2$ kernel values for the pair of points from different classes. Ideally, those points should have a small similarity, preferably zero, for quantum kernels (see, for example, the ideal kernel in Kernel Target Alignment \cite{cristianini2001kernel}). Hence, we can expect that there is a real chance that those kernel values will be estimated by a quantum machine to be vanishing.
The remaining $\frac{m}{2}(\frac{m}{2}-1)$ pairs of points are expected to have non-vanishing similarity. They do not necessarily have to be close to unity, as the kernel matrix can always be rescaled, without any impact on the classification performance. For $m \gg 1$ the number of kernel entries belonging to opposite classes and the same classes will be roughly the same. Therefore, for fidelity quantum kernel, it seems sensible to represent the kernel value set with the median $\kappa_{repr} = median(\mathbb{K})$ and the spread of kernel values as an inter-quartile range $\Delta_{ensemble} = IQR(\mathbb{K})$.

In the case of the concentration avoidance effect for fidelity quantum kernels, we extend the considerations from the above paragraph and propose to exchange the base of the logarithm in formulae (\ref{eq:N_CA_FQ}) and (\ref{eq:N_CA_FQ_F}) to $1-\kappa_{repr}$ and $1-\kappa_{repr,f}$, respectively.
The case of the concentration avoidance effect for projected quantum kernels is much more complicated.
Every value of the projected quantum kernel is a function of $6 N$ measured values, therefore the condition (\ref{eq:PQ_AC}) does not translate easily to the condition on the kernel value and subsequently to the condition of the ensemble of kernel values.
We need to estimate the typical scale, on which measured values differ from the concentration value.
For a single measured value we introduce a convenient notation
\begin{equation}
    \mathcal{M}^{\alpha}_{k}(x_i) = \mu + \varepsilon_k^{\alpha}(x_i),
\end{equation}
where $\alpha \in \{D,R,I\}$ distinguishes the three independent components of the one-qubit density matrix, as in the formula (\ref{eq:projected_rhos}), $k$ numbers the projected qubit registers, while $i$ numbers the data points in the data set $\{x_i\}^m_{i=1}$. 
Observe, that it is the $\varepsilon_k^{\alpha}(x_i)$ that we are interested in the denominator of the formula (\ref{eq:PQ_AC}).

If we have access to the reduced density matrices of the data points, choosing an average norm of the $\varepsilon_k^{\alpha}(x_i)$ for the scale of the typical distance from the concentration value is natural. For example,
\begin{equation}
    \varepsilon^{1}_{\mathcal{R}} = \langle \langle \langle |\varepsilon_k^{\alpha}(x_i)| \rangle_k \rangle_{\alpha} \rangle_x = \frac{1}{3mn} \sum_i \sum_{\alpha} \sum_k |\varepsilon_k^{\alpha}(x_i)|,
\end{equation}
where we have introduced the notation for averages over different variables. $\langle \cdot \rangle_k = \frac{1}{n}\sum_{k} \cdot$ for the average over the projected one-qubit registers, $\langle \cdot \rangle_{\alpha} = \frac{1}{3}\sum_{\alpha} \cdot$ for the average over the independent elements of the one-qubit reduced density matrix, and $\langle \cdot \rangle_{x} = \frac{1}{m}\sum_{i} \cdot$ for the average over the data points.
Then the analogue of the expression (\ref{eq:PQ_AC}) for the whole dataset becomes
\begin{equation}
        \bar{N}_{CA}^{PQ} = \frac{z^2 \mu (1-\mu)}{(\varepsilon^1_{\mathcal{R}})^2}.
\end{equation}
Where we have assumed that, on average, the true measured values are close to the concentration value.

If we have only the access to the projected kernel matrix, we propose a similar approach.
Observe that for the projected quantum kernel, defined in (\ref{eq:projected_kernel}) we have
\begin{equation}\label{eq:logarithm_kappa_PQ}
    -\frac{\ln{[\kappa^{PQ}(x_i,x_j)]}}{6\gamma n} = \langle \langle \left( \varepsilon_k^{\alpha}(x_i) - \varepsilon_k^{\alpha}(x_j) \right)^2 \rangle_k \rangle_{\alpha}.
\end{equation}
We introduce another average $\langle \cdot \rangle_{\mathbb{K}} = \frac{1}{m^2}\sum_i \sum_j$ which is taken over the pairs of data points, hence over the kernel matrix entries.
Applying the above average to the formula (\ref{eq:logarithm_kappa_PQ}), we get
\begin{equation}
    \bigg\langle -\frac{\ln{[\kappa^{PQ}(x_i,x_j)]}}{6\gamma n} \bigg\rangle_{\mathbb{K}} = \langle \langle \langle \left( \varepsilon_k^{\alpha}(x_i) - \varepsilon_k^{\alpha}(x_j) \right)^2 \rangle_k \rangle_{\alpha} \rangle_{\mathbb{K}} = 2 \left( \langle \langle \langle \left( \varepsilon_k^{\alpha}(x_i)\right)^2 \rangle_k \rangle_{\alpha} \rangle_{\mathbb{K}} - \langle \langle \langle \varepsilon_k^{\alpha}(x_i)\varepsilon_k^{\alpha}(x_j)  \rangle_k \rangle_{\alpha} \rangle_{\mathbb{K}} \right).
\end{equation}
Again, we assume that, on average, the true measured values are close to the concentration value, $\langle \varepsilon_k^{\alpha}(x_i) \rangle_{\mathbb{K}}=0$.
Then we can introduce another useful scale, derived solely from the kernel entries,
\begin{equation}
    \varepsilon^{2}_{\mathcal{R}} = \sqrt{\langle \langle \langle \left( \varepsilon_k^{\alpha}(x_i)\right)^2 \rangle_k \rangle_{\alpha} \rangle_{\mathbb{K}}} = \sqrt{- \frac{1}{12 \gamma n} \langle \ln[\kappa^{PQ}(x_i,x_j)] \rangle_{\mathbb{K}} }
\end{equation}
Similarly to, the $\varepsilon^{1}_{\mathcal{R}}$ we can define
\begin{equation}\label{eq:N_CA_PQ_whole}
    \bar{N}_{CA}^{PQ} = \frac{z^2 \mu (1-\mu)}{(\varepsilon^2_{\mathcal{R}})^2}.
\end{equation}

The application of the rules introduced above is discussed in Section \ref{sec:results}, where we use the second definition for the $\tilde{N}_{CA}^{PQ}$ (Eq. (\ref{eq:N_CA_PQ_whole})).
In the view of the exponential value concentration, we determine the behavior of statistical measures of ensembles of kernels as a function of size $n$ of the quantum device. We extrapolate this behavior to the device sizes beyond current capabilities of classical emulation, determine $N$ and assess resources needed for successful quantum kernel estimation. This allows to concretely understand the feasibility of quantum kernel methods for moderate $n$ sizes.

\section{Case Study: Exponential Value Concentration}\label{sec:results}

In this section, we put the material introduced in previous section to use by performing $N$ estimation in the numerical study of the exponential kernel concentration for different kernel families and datasets. First, we introduce a feature map used, then present the numerical results of the exponential value concentration. From the collective behavior of kernel matrix entries, we extrapolate the scaling of statistical measures of kernel values and their spread. The extrapolation is done by fitting the expected exponential function for classically tractable systems. If the fit accuracy exceeds the chosen quality threshold, the behavior for systems beyond our computational capabilities is inferred. In this way, we estimate the minimal number of shots for computing kernel entries for large number of qubits $n$ in different setups.

There are various types of quantum embeddings including, among others, the basis, amplitude, angle, or instantaneous quantum polynomial-time (IQP) embeddings. 
We will focus on the IQP embedding as it leads to a non-separable, hard-to-compute on a classical device quantum state.
In \cite{havlivcek2019supervised}, the authors introduced the ZZ-feature map which is proven to generate states whose inner product is $\#\P$-hard to simulate \cite{goldberg2017complexity}. The single repetition of the ZZ-feature map consists of a Hadamard layer, followed by one- and two-qubit rotations with respect to the data-dependent angle:
\begin{equation}\label{eq:ZZmap}
        \mathcal{U}_{x} = \exp \left( i \sum_{S\subseteq [n]} \phi_S(x) \prod_{i \in S} Z_i \right)H^{\otimes n},
\end{equation}
where $S$ is a set of one- and two-qubit subsets, $Z_i$ is applied on $i^{th}$ qubit, and $\phi_S$ is the angle encoding function. For one-qubit encoding, we have $\phi(x) = x$, while for two qubits, we have $\phi(x,y) = (\pi-x)(\pi-y)$.
The ZZ-feature map consists of $r$ repetitions of the above unitary transformation:
\begin{equation}\label{eq:ZZreps}
    U_x = (\mathcal{U}_{x})^r.
\end{equation}

Increasing the number of repetitions should increase the expressibility of the feature map, moreover $r>1$ prevents from using the classical uniform sampling to estimate the map efficiently \cite{havlivcek2019supervised}.
In Appendix \ref{app:characteristics}, we present an in-depth analysis of the characteristics of the studied feature map.

\begin{figure}[ht!]
    \centering
    
    \begin{subfigure}[b]{\textwidth}
        \centering
        \includegraphics[width=\textwidth]{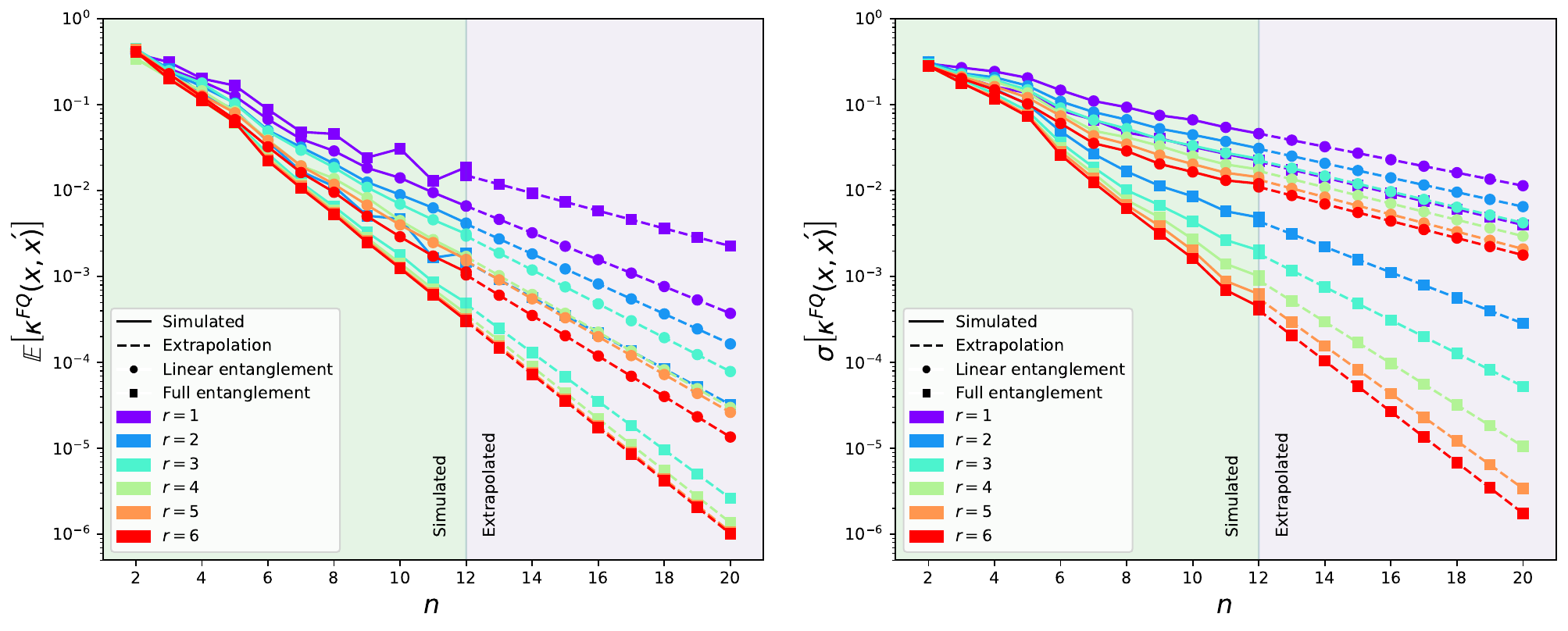}
        \label{fig:plot1}
    \end{subfigure}
    
    \begin{subfigure}[b]{\textwidth}
        \centering
        \includegraphics[width=\textwidth]{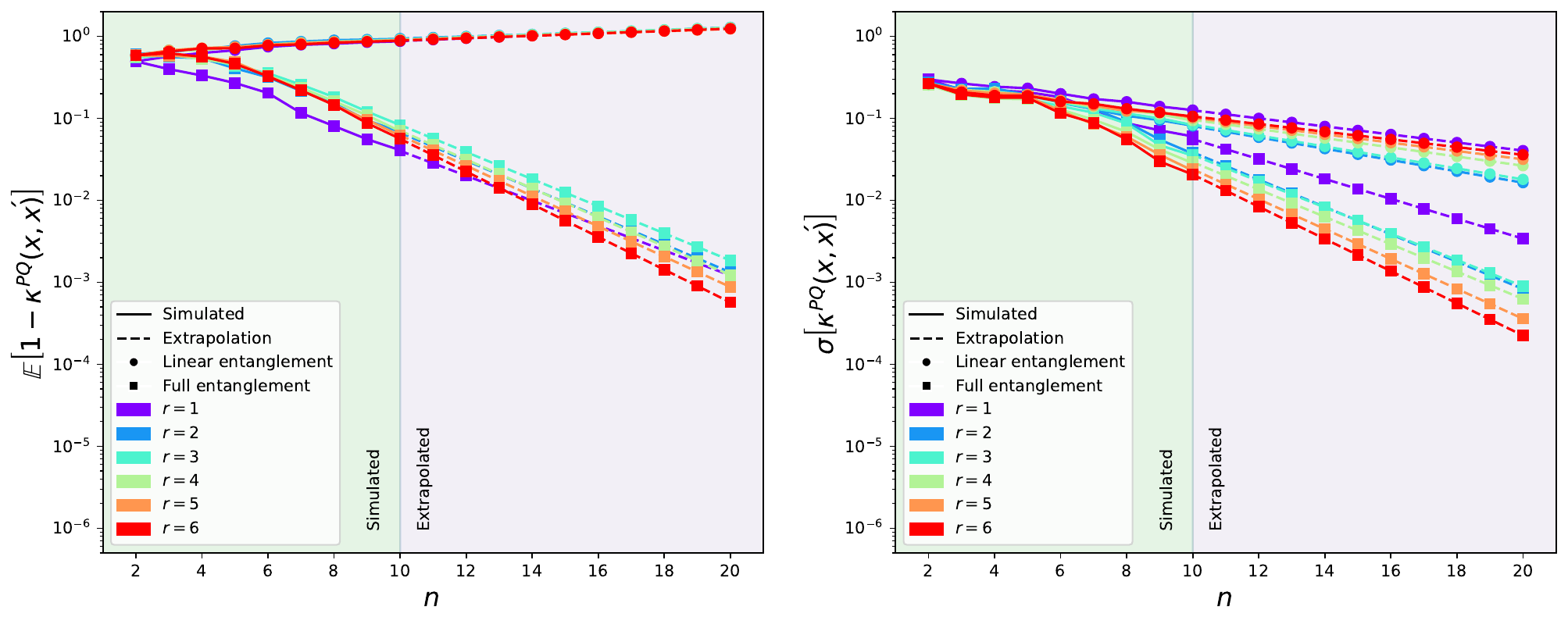}
        \label{fig:plot2}
    \end{subfigure}
    
    \caption{Exponential concentration for the Connectionist Bench dataset, embedded into the quantum state with ZZ-feature map and for the fidelity (Eq.~\ref{eq:fidelity_kernel}) (\textit{upper panel}) and projected kernel (Eq.~\ref{eq:projected_kernel}) (\textit{lower panel}). The dashed lines indicate extrapolations.
    \textit{Left:} Exponential concentration for the mean of independent kernel entries.
    \textit{Right:} Exponential concentration for the standard deviation of independent kernel entries.}
    \label{fig:exo_conc}
\end{figure}

The results for the simulation of the kernel matrix for Connectionist Bench dataset \cite{sonar} are presented in Figure~\ref{fig:exo_conc}. 
They confirm the presence of the exponential value concentration. 
One can see---for the fidelity quantum kernel---that above the size of roughly $8$-qubits, the exponential concentration behavior is an asymptotic one. Therefore, the exponential fit was performed to the results with the first couple points ignored. The decision how many initial points to exclude from the fit was made according to the elbow method with respect to the $R^2$ score of the linear fit to the log-transformed data. This method allows for reliable estimation of the asymptotic behavior, even when the data to which the fit is performed is distributed according to the exponential function with additional sub-exponential factors. Nevertheless, the fit might be impeded by the instability of the results, like in the case of $r=1$ in Figure \ref{fig:exo_conc}, for ``full'' entanglement fidelity kernels.

Now, we turn to the case of the projected quantum kernel. The first observation is that the values for the linear and full entanglement schemes behave differently. In the latter case, we observe the exponential value concentration around the kernel value of $1$. For the linear entanglement, we observe that, on average, independent kernel entries tend to $0$. Upon further investigation it was found that, on average, the $||\rho_k(x)-\rho_k(y)||$ are approximately constant. We remember that with increasing number of qubits $n$ one has more negative terms in the exponential of the projected quantum kernel. Therefore, the supposed concentration of values for the linear entanglement is not confirmed in the range of the tested values of $n$. We exclude this case from the further analysis.



\begin{figure}[ht!]
    \centering
    \includegraphics[width = \textwidth]{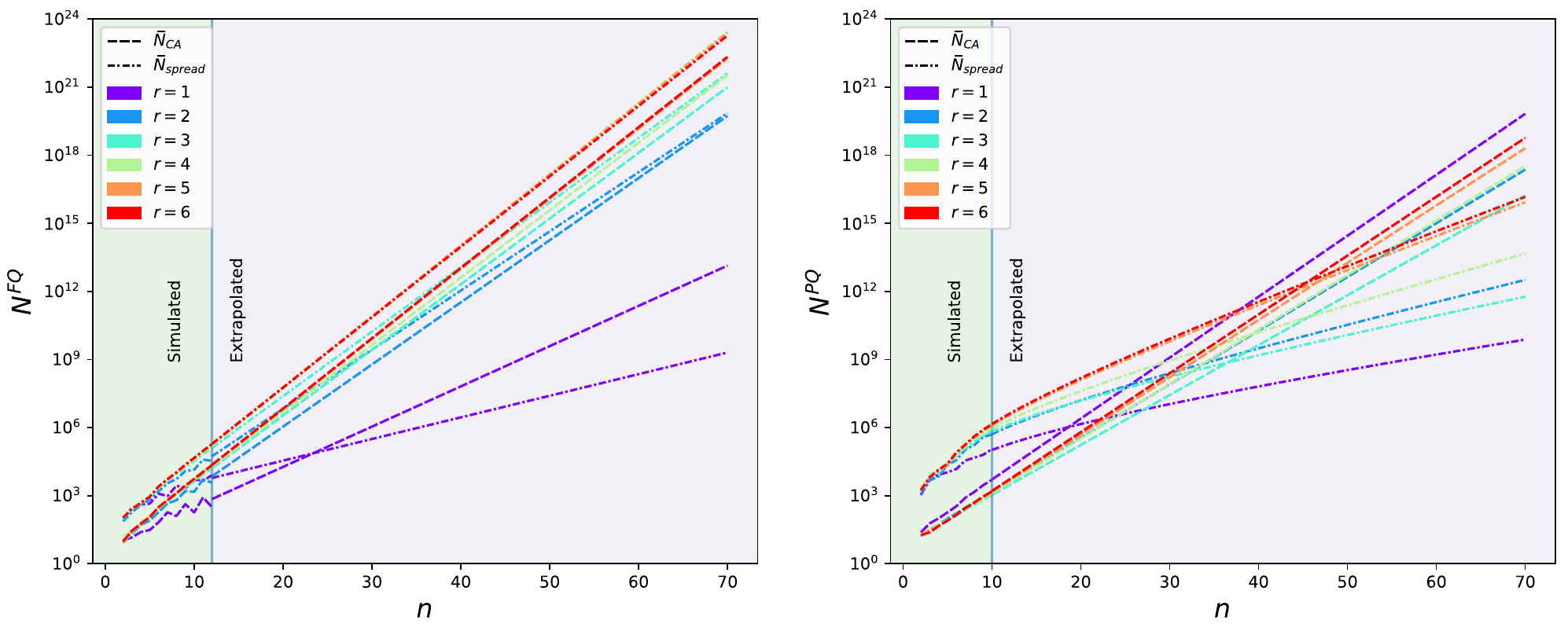} 
    \caption{Number of shots $N$ estimation for Connectionist Bench dataset, embedded into the quantum state with the ZZ-feature map (Eq.~\ref{eq:ZZmap}), for the fidelity (\textit{left}) and projected (\textit{right}) quantum kernel families as a function of number of qubits $n$. The full entanglement strategy was employed in both cases, the results are presented with different feature map repetitions $r$ and are divided into two bound for $N$: spread (Eq.~\ref{eq:Spread}) and concentration avoidance (Eq.~\ref{eq:sr}).}
            \label{fig:N_estimates_sonar}
\end{figure}

In Figure~\ref{fig:N_estimates_sonar}, we present the number of shots $N$ estimates originating from the spread (Eq.~\ref{eq:Spread}) and concentration avoidance (Eq.~\ref{eq:sr}) bounds for the ZZ-feature map (Eq.~\ref{eq:ZZmap}) embedded data points from the Connectionist Bench dataset \cite{sonar}. The estimates are presented for different values of feature map repetitions $r$. The right-hand side of the plots was obtained by extrapolating exponential behavior of the $\tilde{N}_{CA}$ and $\tilde{N}_{spread}$ functions. Observe that for the simulated family of fidelity quantum kernels, the estimated values of the number of shots  $N^{FQ}$ have a similar behavior, with a minor dominance of the spread effect. In the case of the family of projected quantum kernels, we have two distinct regions of dominance of different effects. For small $n$'s, the spread effect clearly dominates, while for large numbers of qubits the concentration effect takes over. 

The estimation of the number of shots is crucial and serves as a basis of quantum experiment planning, resource estimation, and finally the possible utility of quantum kernel methods. In the next section, we discuss the numerical results of the number of shots estimation in the context of computational resource estimation.

\begin{table}[ht!]
\caption{Exponential scaling of the median and inter-quartile range for quantum kernels for different datasets. The $\alpha_X$ values correspond to the $C_X \cdot 2^{\alpha_X n}$ fits of the $X=M$ median ($M_{FQ}$ or $M^{PQ}$) and $X={\rm IQR}$ inter-quartile range for the data. Due to the fact that the exponential concentration was found on an example dataset with fidelity kernels with the \textit{linear} and \textit{full} entanglement strategy, and projected kernels with the \textit{full} strategy only this kernel are considered. The datasets are: sonar---Connectionist Bench dataset~\cite{sonar}; twonorm\cite{twonorm}; indian---the Indian Pines hyperspectral dataset~\cite{PURR1947}. $\times$ indicates fits which did not reach the $R^2$ of $99\%$}
\label{tab:dataset_fits}
\centering
\begin{tabular}{ll|llll|ll|}
  &         & \multicolumn{4}{l|}{Fidelity}                                                      & \multicolumn{2}{l|}{Projected} \\ \cline{3-8} 
  &         & \multicolumn{2}{l|}{Linear}                        & \multicolumn{2}{l|}{Full}     & \multicolumn{2}{l|}{Full}      \\ \hline
r & dataset & $-\alpha_M$ & \multicolumn{1}{l|}{$-\alpha_{IQR}$} & $-\alpha_M$ & $-\alpha_{IQR}$ & $-\alpha_M$  & $-\alpha_{IQR}$ \\ \hline
1 & sonar   & 1.17        & \multicolumn{1}{l|}{1.14}            & 0.30         & 0.42            & 0.31         & 0.66            \\
1 & twonorm & 1.22        & \multicolumn{1}{l|}{1.16}            & $\times$        & $\times$             & 0.24         & 0.36            \\
1 & indian  & 0.72        & \multicolumn{1}{l|}{0.50}             & 0.23        & 0.12            & $\times$          & $\times$             \\ \hline
2 & sonar   & 1.00         & \multicolumn{1}{l|}{0.89}            & 0.69        & 0.70             & 0.15         & 0.55            \\
2 & twonorm & 1.05        & \multicolumn{1}{l|}{0.93}            & $\times$         & $\times$             & $\times$          & 0.4             \\
2 & indian  & 0.77        & \multicolumn{1}{l|}{0.43}            & 0.43        & 0.36            & $\times$          & $\times$             \\ \hline
3 & sonar   & 0.99        & \multicolumn{1}{l|}{0.89}            & 0.96        & 0.95            & 0.11         & 0.49            \\
3 & twonorm & 1.00         & \multicolumn{1}{l|}{0.89}            & 0.96        & 0.95            & $\times$          & 0.37            \\
3 & indian  & 0.86        & \multicolumn{1}{l|}{0.53}            & 0.52        & 0.47            & $\times$          & $\times$             \\ \hline
4 & sonar   & 1.06        & \multicolumn{1}{l|}{0.98}            & 1.01        & 1.01            & $\times$          & 0.58            \\
4 & twonorm & 1.03        & \multicolumn{1}{l|}{0.94}            & 1.01        & 1.00             & $\times$          & 0.42            \\
4 & indian  & 0.89        & \multicolumn{1}{l|}{0.70}             & 0.60         & 0.55            & $\times$          & $\times$             \\ \hline
5 & sonar   & 0.99        & \multicolumn{1}{l|}{0.92}            & 1.02        & 1.03            & 0.13         & 0.64            \\
5 & twonorm & 0.99        & \multicolumn{1}{l|}{0.91}            & 1.02        & 1.02            & 0.08         & 0.49            \\
5 & indian  & 0.93        & \multicolumn{1}{l|}{0.75}            & 0.63        & 0.53            & $\times$          & $\times$             \\ \hline
6 & sonar   & 1.04        & \multicolumn{1}{l|}{0.96}            & 1.03        & 1.03            & 0.17         & 0.66            \\
6 & twonorm & 1.04        & \multicolumn{1}{l|}{0.95}            & 1.03        & 1.02            & 0.13         & 0.55            \\
6 & indian  & 0.96        & \multicolumn{1}{l|}{0.76}            & 0.68        & 0.60             & $\times$          & $\times$             \\
\end{tabular}
\end{table}

\section{Discussion}\label{sec:discussion}

In this work, we presented the idea of using quantum kernels for kernel machines. We reviewed the expressitivity and entanglement characteristics of the quantum embeddings, and introduced two most popular families of quantum kernels: the fidelity and projected quantum kernels. Afterwards, we discussed how the quantum measurement could be understood as a random process and identified the Bernoulli trials for both kernel families. This resulted in the introduction of the system of rules for the number of shots estimation. Then, it was numerically tested for exponentially concentrating kernels to ensure a given level of distinguishability of kernel entries. The method takes into account two effects: the precision of how accurately one can estimate a single kernel entry compared with the kernel values ensemble in the kernel matrix, and the concentration avoidance for the binomial trials.


We focused on the hard-to-compute on a classical device ZZ-feature map, and performed numerical simulations for both families of quantum kernels. As expected, we numerically found that the expressitivity of the quantum kernel depends mostly on how many feature map repetitions $r$ are performed, while it is not strongly influenced by which entanglement strategy is used. It was found, which is in line with the underlying intuition, that the entanglement measure depends strongly on the entanglement strategy of the feature map. For the \textit{linear} entangling strategy, we do not observe an exponential behavior in the simulated number of qubits' range. For the \textit{full} entangling strategy, on the other hand, the behavior is significantly different, and the measure of entanglement behaves exponentially.

In the available range of simulated qubit numbers $n$, we identified the exponential kernel concentration for all entanglement strategies for the fidelity quantum kernel, and for the \textit{full} entanglement strategy for the projected quantum kernel. The severity of the value concentration was tested for different hyperparameters of the ZZ-feature map, with different numbers of repetitions $r$ and different entanglement strategies. 
In general, we confirm that with the greater expressitivity and entanglement the effect of exponential value concentration is more severe. 
The fidelity quantum kernel is more sensitive to the growth of expressibility, while the exponential concentration of the projected quantum kernel is influenced mostly by an entanglement strategy.

We used the introduced method for estimating the number of minimal circuit repetitions to the simulated data. The threshold number of qubits above which the stable exponential growth of the minimal circuit repetitions occurred was identified, and the corresponding growth rates were found.
The minimal number of circuit repetitions needed for the kernel values distinguishability can be translated to the runtime. This can be done by multiplying the number of shots by the number of gate layers by their execution time and also adding a measurement time for each shot. The energy consumption can be estimated by multiplying the expected runtime by the expected power of a single physical qubit which is roughly estimated to be around $30$\,mW (see Appendix~\ref{App:error_estimates}). 

In order to give an additional meaning of the expected runtime of the quantum device, we performed an estimate on the classical runtime behavior for the quantum kernel calculation. 
This analysis was performed by calculating quantum kernel entries on a classical device with the \texttt{qiskit} software, and the number of performed floating point operations was recorded. 
Classical computers have been in development for much longer than quantum computers, resulting in significantly more advanced technology and software. This is evident in their faster command execution, greater parallelization capabilities, optimized calculations, and minimal error rates.
The growth factors $\alpha^C_{FQ/PQ}$ (defined analogously to $\alpha_X$'s in Table \ref{tab:dataset_fits}) of the runtime of a classical device were found to be $\alpha^C_{FQ} \sim 1.07$ and $\alpha^C_{PQ} \sim 2.30$.
Additionally, it must be kept in mind that other than the direct computation methods can be found to obtain quantum kernel entries on a classical device. They can use the fact that kernel entries have to be estimated with similar or smaller errors that on the quantum device, which could greatly improve the scaling of the classical runtime.
Nevertheless, we can treat $\alpha^C_X$'s as a maximal bound on the usefulness of quantum kernels. 
We assume that the estimated quantum kernel entries lead to the superior classification performance, hence to obtain similar performance on a classical device one has to simulate the same kernel entries on a classical device. Additionally, we exclude the possibility that the same classification performance can be obtained with simpler, more simulable quantum kernels \cite{slattery2023numerical}.
In Table \ref{tab:dataset_fits}, we observe that the family of fidelity quantum kernels have a scaling of runtime close to $\alpha^C_{FQ}$. Therefore, one can expect this quantum kernel family will have significantly reduced chances for utility.
On the other hand, projected quantum kernels estimated on a quantum device are expected to have much better asymptotic scaling behavior of runtime compared to the use of a classical device.

\begin{figure}[ht!]
    \centering
    
        \centering
        \includegraphics[width=\textwidth]{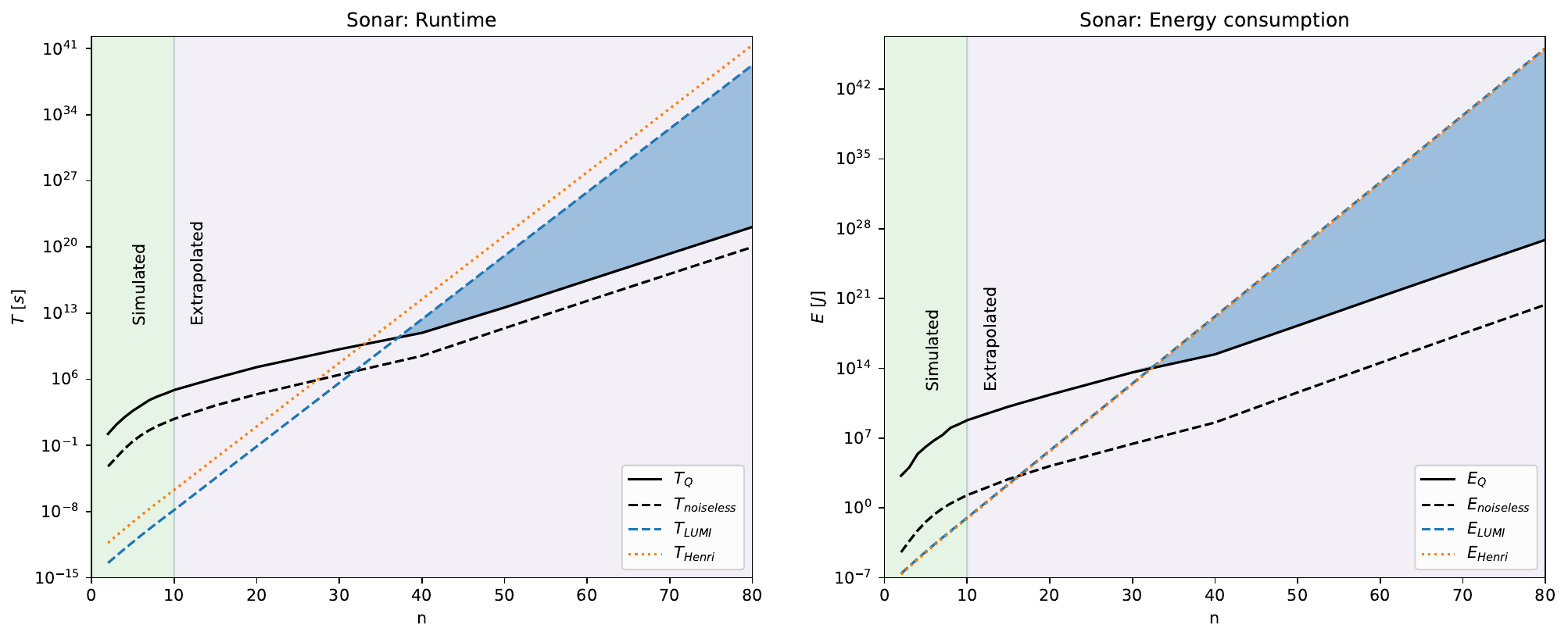}
        \label{fig:plot1}
    
%
    \caption{The quantum and classical estimation of runtime (\textit{left}) and energy consumption (\textit{right}) for obtaining a single kernel value. The blue region indicates the possible benefit of using quantum device for faster or more energy efficient error-corrected computation. $T_Q$ ($E_Q$), $T_{noiseless}$ ($E_{noiseless}$), $T_{LUMI}$ ($E_{LUMI}$), and $T_{Henri}$ ($E_{Henri}$) represent the estimated computational time (energy consumption) for the error-corrected quantum device, the noiseless quantum device, the LUMI supercomputer, and the Henri supercomputer, respectively. 
    Resource estimations are performed for projected quantum kernels for the Connectionist Bench dataset used throughout the paper and embedded with a ZZ-feature map (Eq.~\ref{eq:ZZmap}) with $r=6$. 
    }
    \label{fig:QRE}
\end{figure}
In Figure~\ref{fig:QRE}, we compare the expected runtime and energy consumption for classical and quantum devices for projected quantum kernels. The left (right) plots show the runtime (energy consumption) estimation.
The resource utilization for the classical simulation was calculated by taking the estimated number of floating point operations and translating them to the runtime and energy consumption of two supercomputers: LUMI (rank 5/rank 7 on TOP500/Green500 November 2023 list \url{www.top500.org}), which is currently highly developed computational infrastructure for HPC-quantum integration and Henri (rank 293/rank 1), which is the leading computational device on the Green500 list.
The results of the quantum resources' utilization are presented both for the idealized, noiseless machines ($T_{noiseless}/E_{noiseless}$) and machines with implemented error correction scheme ($T_{Q}/E_{Q}$). 
The parameters of error-corrected quantum circuits are obtained with the use of \texttt{Azure Quantum Resource Estimator}.
The blue-filled area indicates the expected quantum utility within the above assumptions.
In general, we expect quantum utility for roughly $n>40$ in all the presented cases.
We have found, that
the possible utility is achieved with, at least, $n=37$. The alarming result is that even for this size of a quantum device the single data point tomography time is of order of $\approx 10^{10}s$, which is clearly prohibitive for any useful computation. The quantum utility in energy consumption in general is possibly achieved for smaller number of qubits, in the case of Connectionist Bench example for $n=32$.

In our opinion, this casts an even darker shadow on the quantum kernel methods as we understand them today.
In the work \cite{thanasilp2022exponential}, we learned that there is no computational complexity separation between the classical simulation of quantum kernels and quantum kernel estimation for general real-life problems.
In our study, we strengthen this result by numerical simulations that point out that even for moderate problems sizes the possible quantum speedups (or energy savings) tend to appear at the systems' sizes which are already hard to use in real life.

We are aware of the limitations of the performed numerical analysis. 
It was performed only for the binary classification task, for a limited number of datasets. 
The estimates of the classical simulation runtime is highly pessimistic, as it only takes into account number of floating point operations for perfect simulation performed with general-purpose software, which was not specifically optimized for the task.
Nevertheless, the numerical study demonstrates how the theoretical ideas introduced in this work can be used for particular problems. The slightly underestimated number of qubits, beyond which we should be able to find the utility of the QKE, in general agrees with the expected sizes of the useful Noisy, Intermediate-Scale Quantum devices \cite{preskill2018quantum}. Although, already in the few studied datasets we observe the dependence of the exponential value concentration behavior on the data, we believe that the conclusion of prohibitive runtime for the useful QKE procedure remains robust.

\subsection{Do Those Results Exclude Completely the Quantum Kernel Utility?}

We wish to emphasize that, with the current understanding of quantum kernel methods, the landscape of possible advantageous use cases shrinks, but the utility is not yet completely excluded:
\begin{itemize}
    \item There are claims that some of quantum kernel families avoid the exponential concentration issue \cite{suzuki2022quantum}. Yet it is conjectured that quantum systems which avoid concentration problems exploit symmetries in the studied systems \cite{cerezo2023does}. If this is the case, classical methods might also utilize the symmetries for faster simulations, and the potential quantum advantage would be harder to achieve. 
    \item Another important observation is that the severity of the exponential value concentration inexorably depends on the structure of the data used, see Table \ref{tab:dataset_fits}. Interestingly, the hyperspectral dataset of Indian Pines \cite{PURR1947}, exhibits a much milder exponential value concentration effect (with the same feature map hyperparameters) than other considered datasets. The explanation of what kind of data decreases severity of the exponential value concentration is an interesting research direction, that goes beyond the scope of the work presented here.
    \item For the linear entanglement strategy, we have not found a strong exponential concentration effect (see Figure~\ref{fig:exo_conc}) in the simulated range of quantum device sizes. Either the problem of the exponential concentration starts to be visible for higher (than simulated) number of qubits, or there is a critical amount of entanglement below which the problem is non-existent. Keep in mind that the critical amount of entanglement might still allow for efficient classical simulation of the system \cite{vidal2003efficient}.
\end{itemize}


\section{Conclusions}\label{sec:conclusions}

In recent years, significant research efforts have been directed toward comprehending the characteristics and performance capabilities of quantum kernel methods.
This primarily involves attempts to enhance metrics strictly dedicated to the machine learning task-oriented performance, such as classification accuracy.
However, our approach in this study diverges from the mainstream. 
We begin with the assumption that an advantageous machine learning performance can indeed be achieved with quantum kernel methods and seek to answer the question whether it is more beneficial to run the actual quantum kernel estimation rather than to perform the kernel simulation on a classical device.
To this end, we have introduced the methods for estimating the number of shots required for a successful quantum kernel estimation, which consequently enables us to evaluate the expected resources needed for obtaining quantum kernels. 
During the estimation of the number of shots, we identify two main effects that must be taken into account and exhibit distinct behaviors: spread and concentration. 
Through numerical experiments, we validated the severity of exponential quantum concentration, assessed the efficacy of the introduced methods, and conducted a preliminary comparison of classical and quantum resources required for obtaining quantum kernels. 
The results indicate the potential for quantum speedups or energy savings in real-life problems. 
However, the actual runtime of the quantum routines renders them impractical, even for moderate-sized problems. 
Consequently, we may conclude that the versatility and utility of quantum kernel methods may not be as substantial as initially anticipated. 
Therefore, a significantly greater research effort must be undertaken to comprehend the actual capabilities and limitations of these methods before making claims regarding their potential utility in real-life problems.

\section*{Comment on Related Work}
In our initial planning, we aimed to extend the work presented in the preprint of \cite{thanasilp2022exponential}. However, it has come to our attention that this preprint has recently been published, and it underwent substantial expansions during the peer review process. Notably, the published version partially overlaps with the contributions we present in this paper. It is important to note that our research was conducted simultaneously and independently of these recent developments.

\section*{Acknowledgements}

We warmly thanks Kenton Barnes for useful discussions regarding the energy requirements of surface code decoders, Alistair Francis and Peter Naylor for the discussions regarding concentration avoidance effect. This work was supported by the European Union's Horizon Europe Research and Innovation programme under the Marie Sklodowska-Curie Actions \& Support to Experts programme (MSCA Postdoctoral Fellowships) - grant agreement No. 101108284. This work was funded by the European Space Agency, and supported by the ESA $\Phi$-lab (\url{https://philab.phi.esa.int/}) AI-enhanced Quantum Computing for Earth Observation (QC4EO) initiative, under ESA contract No. 4000137725/22/NL/GLC/my. JN was supported by the Silesian University of Technology grant for maintaining and developing research potential.

\bibliography{bibliography}

\begin{thebibliography}{10}

\bibitem{biamonte2017quantum}
J.~Biamonte, P.~Wittek, N.~Pancotti, P.~Rebentrost, N.~Wiebe, and S.~Lloyd,
  ``Quantum machine learning,'' {\em Nature}, vol.~549, no.~7671, pp.~195--202,
  2017.

\bibitem{schuld2015introduction}
M.~Schuld, I.~Sinayskiy, and F.~Petruccione, ``An introduction to quantum
  machine learning,'' {\em Contemporary Physics}, vol.~56, no.~2, pp.~172--185,
  2015.

\bibitem{havlivcek2019supervised}
V.~Havl{\'\i}{\v{c}}ek, A.~D. C{\'o}rcoles, K.~Temme, A.~W. Harrow, A.~Kandala,
  J.~M. Chow, and J.~M. Gambetta, ``Supervised learning with quantum-enhanced
  feature spaces,'' {\em Nature}, vol.~567, no.~7747, pp.~209--212, 2019.

\bibitem{schuld2019quantum}
M.~Schuld and N.~Killoran, ``Quantum machine learning in feature hilbert
  spaces,'' {\em Physical review letters}, vol.~122, no.~4, p.~040504, 2019.

\bibitem{miyabe2023quantum}
S.~Miyabe, B.~Quanz, N.~Shimada, A.~Mitra, T.~Yamamoto, V.~Rastunkov,
  D.~Alevras, M.~Metcalf, D.~J. King, M.~Mamouei, {\em et~al.}, ``Quantum
  multiple kernel learning in financial classification tasks,'' {\em arXiv
  preprint arXiv:2312.00260}, 2023.

\bibitem{grossi2022mixed}
M.~Grossi, N.~Ibrahim, V.~Radescu, R.~Loredo, K.~Voigt, C.~Von~Altrock, and
  A.~Rudnik, ``Mixed quantum--classical method for fraud detection with quantum
  feature selection,'' {\em IEEE Transactions on Quantum Engineering}, vol.~3,
  pp.~1--12, 2022.

\bibitem{delilbasic2021quantum}
A.~Delilbasic, G.~Cavallaro, M.~Willsch, F.~Melgani, M.~Riedel, and
  K.~Michielsen, ``Quantum support vector machine algorithms for remote sensing
  data classification,'' in {\em 2021 IEEE International Geoscience and Remote
  Sensing Symposium IGARSS}, pp.~2608--2611, IEEE, 2021.

\bibitem{miroszewski2023detecting}
A.~Miroszewski, J.~Mielczarek, G.~Czelusta, F.~Szczepanek, B.~Grabowski,
  B.~Le~Saux, and J.~Nalepa, ``Detecting clouds in multispectral satellite
  images using quantum-kernel support vector machines,'' {\em IEEE Journal of
  Selected Topics in Applied Earth Observations and Remote Sensing}, vol.~16,
  pp.~7601--7613, 2023.

\bibitem{cao2018potential}
Y.~Cao, J.~Romero, and A.~Aspuru-Guzik, ``Potential of quantum computing for
  drug discovery,'' {\em IBM Journal of Research and Development}, vol.~62,
  no.~6, pp.~6--1, 2018.

\bibitem{chan2021sissa}
J.~Chan, C.~Zhou, M.~Livny, S.~Yoo, F.~Carminati, W.~Guan, A.~Wang,
  P.~Spentzouris, S.~Sun, A.~C. Li, {\em et~al.}, ``Sissa: Application of
  quantum machine learning to high energy physics analysis at lhc using ibm
  quantum computer simulators and ibm quantum computer hardware,'' {\em
  Proceedings of Science}, p.~930, 2021.

\bibitem{liu2021rigorous}
Y.~Liu, S.~Arunachalam, and K.~Temme, ``A rigorous and robust quantum speed-up
  in supervised machine learning,'' {\em Nature Physics}, vol.~17, no.~9,
  pp.~1013--1017, 2021.

\bibitem{vasques2023application}
X.~Vasques, H.~Paik, and L.~Cif, ``Application of quantum machine learning
  using quantum kernel algorithms on multiclass neuron m-type classification,''
  {\em Scientific Reports}, vol.~13, no.~1, p.~11541, 2023.

\bibitem{mengoni2019kernel}
R.~Mengoni and A.~Di~Pierro, ``Kernel methods in quantum machine learning,''
  {\em Quantum Machine Intelligence}, vol.~1, no.~3-4, pp.~65--71, 2019.

\bibitem{herrmann2023quantum}
N.~Herrmann, D.~Arya, M.~W. Doherty, A.~Mingare, J.~C. Pillay, F.~Preis, and
  S.~Prestel, ``Quantum utility--definition and assessment of a practical
  quantum advantage,'' {\em arXiv preprint arXiv:2303.02138}, 2023.

\bibitem{hibat2023framework}
M.~Hibat-Allah, M.~Mauri, J.~Carrasquilla, and A.~Perdomo-Ortiz, ``A framework
  for demonstrating practical quantum advantage: Racing quantum against
  classical generative models,'' {\em arXiv preprint arXiv:2303.15626}, 2023.

\bibitem{thanasilp2022exponential}
S.~Thanasilp, S.~Wang, M.~Cerezo, and Z.~Holmes, ``Exponential concentration in
  quantum kernel methods,'' {\em Nature Communications}, vol.~15, no.~1,
  p.~5200, 2024.

\bibitem{huang2021power}
H.-Y. Huang, M.~Broughton, M.~Mohseni, R.~Babbush, S.~Boixo, H.~Neven, and
  J.~R. McClean, ``Power of data in quantum machine learning,'' {\em Nature
  communications}, vol.~12, no.~1, p.~2631, 2021.

\bibitem{suzuki2022quantum}
Y.~Suzuki, H.~Kawaguchi, and N.~Yamamoto, ``Quantum fisher kernel for
  mitigating the vanishing similarity issue,'' {\em Quantum Science and
  Technology}, 2022.

\bibitem{cerezo2023does}
M.~Cerezo, M.~Larocca, D.~Garc{\'\i}a-Mart{\'\i}n, N.~Diaz, P.~Braccia,
  E.~Fontana, M.~S. Rudolph, P.~Bermejo, A.~Ijaz, S.~Thanasilp, {\em et~al.},
  ``Does provable absence of barren plateaus imply classical simulability? or,
  why we need to rethink variational quantum computing,'' {\em arXiv preprint
  arXiv:2312.09121}, 2023.

\bibitem{sonar}
T.~Sejnowski and R.~Gorman, ``{Connectionist Bench (Sonar, Mines vs. Rocks)}.''
  UCI Machine Learning Repository.
\newblock {DOI}: https://doi.org/10.24432/C5T01Q.

\bibitem{cortes1995support}
C.~Cortes and V.~Vapnik, ``Support-vector networks,'' {\em Machine learning},
  vol.~20, pp.~273--297, 1995.

\bibitem{smola2004tutorial}
A.~J. Smola and B.~Sch{\"o}lkopf, ``A tutorial on support vector regression,''
  {\em Statistics and computing}, vol.~14, pp.~199--222, 2004.

\bibitem{scholkopf1998nonlinear}
B.~Sch{\"o}lkopf, A.~Smola, and K.-R. M{\"u}ller, ``Nonlinear component
  analysis as a kernel eigenvalue problem,'' {\em Neural computation}, vol.~10,
  no.~5, pp.~1299--1319, 1998.

\bibitem{berlinet2011reproducing}
A.~Berlinet and C.~Thomas-Agnan, {\em Reproducing kernel Hilbert spaces in
  probability and statistics}, vol.~XII, 355 pages.
\newblock Springer Science \& Business Media, 1~ed., June 2011.

\bibitem{larson2019review}
N.~B. Larson, J.~Chen, and D.~J. Schaid, ``A review of kernel methods for
  genetic association studies,'' {\em Genetic epidemiology}, vol.~43, no.~2,
  pp.~122--136, 2019.

\bibitem{hofmann2008kernel}
T.~Hofmann, B.~Sch{\"o}lkopf, and A.~J. Smola, ``{Kernel methods in machine
  learning},'' {\em The Annals of Statistics}, vol.~36, no.~3, pp.~1171 --
  1220, 2008.

\bibitem{apsemidis2020review}
A.~Apsemidis, S.~Psarakis, and J.~M. Moguerza, ``A review of machine learning
  kernel methods in statistical process monitoring,'' {\em Computers \&
  Industrial Engineering}, vol.~142, p.~106376, 2020.

\bibitem{preskill2018quantum}
J.~Preskill, ``Quantum computing in the nisq era and beyond,'' {\em Quantum},
  vol.~2, p.~79, 2018.

\bibitem{goldberg2017complexity}
L.~A. Goldberg and H.~Guo, ``The complexity of approximating complex-valued
  ising and tutte partition functions,'' {\em computational complexity},
  vol.~26, pp.~765--833, 2017.

\bibitem{hubregtsen2022training}
T.~Hubregtsen, D.~Wierichs, E.~Gil-Fuster, P.-J.~H. Derks, P.~K. Faehrmann, and
  J.~J. Meyer, ``Training quantum embedding kernels on near-term quantum
  computers,'' {\em Physical Review A}, vol.~106, no.~4, p.~042431, 2022.

\bibitem{d2003quantum}
G.~M. D'Ariano, M.~G. Paris, and M.~F. Sacchi, ``Quantum tomography,'' {\em
  Advances in imaging and electron physics}, vol.~128, no.~10.1016,
  pp.~S1076--5670, 2003.

\bibitem{gosset2018compressed}
D.~Gosset and J.~Smolin, ``A compressed classical description of quantum
  states,'' {\em arXiv preprint arXiv:1801.05721}, 2018.

\bibitem{aaronson2018shadow}
S.~Aaronson, ``Shadow tomography of quantum states,'' in {\em Proceedings of
  the 50th annual ACM SIGACT symposium on theory of computing}, pp.~325--338,
  2018.

\bibitem{cristianini2001kernel}
N.~Cristianini, J.~Shawe-Taylor, A.~Elisseeff, and J.~Kandola, ``On
  kernel-target alignment,'' {\em Advances in neural information processing
  systems}, vol.~14, 2001.

\bibitem{twonorm}
``Twonorm dataset.''
  \url{https://www.cs.toronto.edu/~delve/data/twonorm/desc.html}.

\bibitem{PURR1947}
M.~F. Baumgardner, L.~L. Biehl, and D.~A. Landgrebe, ``220 band aviris
  hyperspectral image data set: June 12, 1992 indian pine test site 3,'' {\em
  Purdue University Research Repository}, vol.~10, no.~7, p.~991, 2015.

\bibitem{slattery2023numerical}
L.~Slattery, R.~Shaydulin, S.~Chakrabarti, M.~Pistoia, S.~Khairy, and S.~M.
  Wild, ``Numerical evidence against advantage with quantum fidelity kernels on
  classical data,'' {\em Physical Review A}, vol.~107, no.~6, p.~062417, 2023.

\bibitem{vidal2003efficient}
G.~Vidal, ``Efficient classical simulation of slightly entangled quantum
  computations,'' {\em Physical review letters}, vol.~91, no.~14, p.~147902,
  2003.

\bibitem{holmes2022connecting}
Z.~Holmes, K.~Sharma, M.~Cerezo, and P.~J. Coles, ``Connecting ansatz
  expressibility to gradient magnitudes and barren plateaus,'' {\em PRX
  Quantum}, vol.~3, no.~1, p.~010313, 2022.

\bibitem{nakaji2021expressibility}
K.~Nakaji and N.~Yamamoto, ``Expressibility of the alternating layered ansatz
  for quantum computation,'' {\em Quantum}, vol.~5, p.~434, 2021.

\bibitem{sim2019expressibility}
S.~Sim, P.~D. Johnson, and A.~Aspuru-Guzik, ``Expressibility and entangling
  capability of parameterized quantum circuits for hybrid quantum-classical
  algorithms,'' {\em Advanced Quantum Technologies}, vol.~2, no.~12,
  p.~1900070, 2019.

\bibitem{ali2020chaos}
T.~Ali, A.~Bhattacharyya, S.~S. Haque, E.~H. Kim, N.~Moynihan, and J.~Murugan,
  ``Chaos and complexity in quantum mechanics,'' {\em Physical Review D},
  vol.~101, no.~2, p.~026021, 2020.

\bibitem{Fowler2012Sep}
A.~G. Fowler, M.~Mariantoni, J.~M. Martinis, and A.~N. Cleland, ``{Surface
  codes: Towards practical large-scale quantum computation},'' {\em Phys. Rev.
  A}, vol.~86, p.~032324, Sept. 2012.

\bibitem{Terhal2015Apr}
B.~M. Terhal, ``{Quantum error correction for quantum memories},'' {\em Rev.
  Mod. Phys.}, vol.~87, pp.~307--346, Apr. 2015.

\bibitem{beverland2022assessing}
M.~E. Beverland, P.~Murali, M.~Troyer, K.~M. Svore, T.~Hoefler, V.~Kliuchnikov,
  G.~H. Low, M.~Soeken, A.~Sundaram, and A.~Vaschillo, ``Assessing requirements
  to scale to practical quantum advantage,'' {\em arXiv preprint
  arXiv:2211.07629}, 2022.

\bibitem{Fellous-Asiani2023Oct}
M.~Fellous-Asiani, J.~H. Chai, Y.~Thonnart, H.~K. Ng, R.~S. Whitney, and
  A.~Auff{\ifmmode\grave{e}\else\`{e}\fi}ves, ``{Optimizing Resource
  Efficiencies for Scalable Full-Stack Quantum Computers},'' {\em PRX Quantum},
  vol.~4, p.~040319, Oct. 2023.

\bibitem{Parma2014}
V.~Parma, ``{Cryostat Design},'' in {\em Superconductivity for Accelerators}
  (R.~Bailey, ed.), (CERN Yelow Reports: School Proceedings), pp.~pp353--399,
  CERN, Geneva, 2014.
\newblock \href{arXiv:1501.07154}{arXiv:1501.07154}.

\bibitem{martin2022energy}
M.~J. Martin, C.~Hughes, G.~Moreno, E.~B. Jones, D.~Sickinger, S.~Narumanchi,
  and R.~Grout, ``Energy use in quantum data centers: Scaling the impact of
  computer architecture, qubit performance, size, and thermal parameters,''
  {\em IEEE Transactions on Sustainable Computing}, vol.~7, no.~4,
  pp.~864--874, 2022.

\bibitem{Park2021Feb}
J.-S. Park, S.~Subramanian, L.~Lampert, T.~Mladenov, I.~Klotchkov, D.~J.
  Kurian, E.~Juarez-Hernandez, B.~Perez-Esparza, S.~R. Kale, K.~T.~A. Beevi,
  S.~Premaratne, T.~Watson, S.~Suzuki, M.~Rahman, J.~B. Timbadiya, S.~Soni, and
  S.~Pellerano, ``{13.1 A Fully Integrated Cryo-CMOS SoC for Qubit Control in
  Quantum Computers Capable of State Manipulation, Readout and High-Speed Gate
  Pulsing of Spin Qubits in Intel 22nm FFL FinFET Technology},'' in {\em {2021
  IEEE International Solid- State Circuits Conference (ISSCC)}}, vol.~64,
  pp.~208--210, IEEE, Feb. 2021.

\bibitem{frank2022cryo}
D.~J. Frank, S.~Chakraborty, K.~Tien, P.~Rosno, T.~Fox, M.~Yeck, J.~A. Glick,
  R.~Robertazzi, R.~Richetta, J.~F. Bulzacchelli, {\em et~al.}, ``A cryo-cmos
  low-power semi-autonomous qubit state controller in 14nm finfet technology,''
  in {\em 2022 IEEE International Solid-State Circuits Conference (ISSCC)},
  vol.~65, pp.~360--362, IEEE, 2022.

\bibitem{Kang2022Aug}
K.~Kang, D.~Minn, S.~Bae, J.~Lee, S.~Kang, M.~Lee, H.-J. Song, and J.-Y. Sim,
  ``{A 40-nm Cryo-CMOS Quantum Controller IC for Superconducting Qubit},'' {\em
  IEEE Journal of Solid-State Circuits}, pp.~1--14, Aug. 2022.

\bibitem{barber2023real}
B.~Barber, K.~M. Barnes, T.~Bialas, O.~Bu{\u{g}}dayc{\i}, E.~T. Campbell, N.~I.
  Gillespie, K.~Johar, R.~Rajan, A.~W. Richardson, L.~Skoric, {\em et~al.}, ``A
  real-time, scalable, fast and highly resource efficient decoder for a quantum
  computer,'' {\em arXiv preprint arXiv:2309.05558}, 2023.

\bibitem{Kenton_Barnes}
{{Kenton Barnes, Riverlane}}. Private communication, email.

\bibitem{horsman2012surface}
D.~Horsman, A.~G. Fowler, S.~Devitt, and R.~Van~Meter, ``Surface code quantum
  computing by lattice surgery,'' {\em New Journal of Physics}, vol.~14,
  no.~12, p.~123011, 2012.

\bibitem{litinski2019game}
D.~Litinski, ``A game of surface codes: Large-scale quantum computing with
  lattice surgery,'' {\em Quantum}, vol.~3, p.~128, 2019.

\bibitem{delfosse2021almost}
N.~Delfosse and N.~H. Nickerson, ``Almost-linear time decoding algorithm for
  topological codes,'' {\em Quantum}, vol.~5, p.~595, 2021.

\bibitem{github}
A.~Miroszewski, ``Isoqaetnosiqkm,
  \url{https://github.com/Quantum-Cosmos-Lab/ISoQAEtNoSiQKM},'' 2024.

\end{thebibliography}
\bibliographystyle{ieeetr}

\newpage

\appendix






\section{Table of abbreviations}\label{app:abbreviations}

\begin{table}[h!]
\caption{Significance of various abbreviations, symbols, and acronyms used throughout the paper.}
\label{tab:abbreviations}
\begin{center}
\begin{tabular}{ll}
\hline
Abbreviation & Meaning \\ \hline
QML                     & Quantum machine learning       \\
QKE                     & Quantum kernel estimation       \\
$N$                     & Number of shots, number of circuit runs\\ 
$\tilde{N}_{spread/CA}$ & Sufficient number of shots to satisfy bound for spread/concentration avoidance effects\\
$\tilde{N}$             & $max(N_{spread}, N_{CA})$ \\
$\bar{N}_{spread/CA}$   & Representative $N_{spread/CA}$ for the whole kernel matrix \\
$\bar{N}$             & $max(\bar{N}_{spread}, \bar{N}_{CA})$ \\
$n$                     & Number of qubits\\ 
$m$                     & Number of classical data points in the dataset\\ 
$x$                     & Classical datum\\ 
$\kappa^{FQ/PQ}$        & True value of the fidelity/projected quantum kernel\\ 
$\hat{\kappa}^{FQ/PQ}$  & Measured value of the fidelity/projected quantum kernel, a random variable\\ 
$\rho_k$                & Reduced density matrix with all, except $k^{th}$, qubit registers traced out\\
$\hat{\mathcal{M}}$     & Measured value during the QKE, a random variable\\
                        & Used when the kernel value is a non-linear function of the measurement outcome\\
$\mathbb{E}[\hat{\mathcal{M}}]$           & True value of the measured value\\
$\mu_{\mathcal{M}}$     & Concentration value corresponding to the measured value $\hat{\mathcal{M}}$\\ 
IQR                     & Inter-quartile range \\
$\rho_f, \mathcal{M}^f, N^f,\dots$           & $f$ sub/super-script indicated that we are considering noisy computation \\ 
$\Delta_{kernel\ value}$      & Uncertainty related to the measurement of the kernel value\\
$\Delta_{ensemble}$           & Measure of the spread of the independent kernel values in the Gram matrix\\
$C_X$                   & Initial value constant in the exponential fitting formula $C_X \cdot 2^{\alpha_X n}$\\
$\alpha_X$                   & Growth rate constant in the exponential fitting formula $C_X \cdot 2^{\alpha_X n}$\\
\hline
\end{tabular}
\end{center}
\end{table}







\section{Characteristics of the Feature Map}\label{app:characteristics}
\subsection{Expressibility}

Generally speaking, expressibility connects with the size of the hypotheses set, from which machine learning algorithms can learn. 
Ansatze with high expressibility can embed data on almost all available states, while low expressibility ansatze explore only a small subset of available directions in the target space. 
In QML, expressibility is customarily measured by comparing the distribution of the embedded data with Haar random state distribution on the target Hilbert space \cite{holmes2022connecting,nakaji2021expressibility, sim2019expressibility}.

The expressibility of the circuit is often estimated in terms of the following superoperator \cite{sim2019expressibility, nakaji2021expressibility}:
\begin{equation}\label{eq:A_superoperator}
    \mathcal{A}^{(t)}_{\mathbb{U}} (\cdot) := \int_{Haar} d\mu(V) V^{\otimes t}(\cdot) (V^{\dagger})^{\otimes t} - \int_{\mathbb{U}} dU U^{\otimes t} (\cdot) (U^{\dagger})^{\otimes t},
\end{equation}
where $d\mu(V)$ is the volume element of the Haar measure and $dU$ is the volume element over the ansatz.
The useful measure stemming from the above superoperator is defined as a following trace norm
\begin{equation}\label{eq:general_trace_norm}
    \varepsilon^{(t)}_{\mathbb{U}, p}(\rho_0) =  || \mathcal{A}^{(t)}_{\mathbb{U}}(\rho_0) ||_p = \Tr^{\frac{1}{p}}\left[ |\mathcal{A}^{(t)}_{\mathbb{U}}(\rho_0)|^p\right],
\end{equation}
where $\rho_0$ is the initial state and $|T| = \sqrt{T^{\dagger}T}$. 
In the considerations about quantum kernels, it is sufficient to consider $t=2$, then $\varepsilon^{(2)}_{\mathbb{U}, p}(\rho_0)$ measures how far $\mathbb{U}$ is from forming a 2-design. In the case of $\varepsilon^{(2)}_{\mathbb{U}, p}(\rho_0)=0$ averaging over elements of $\mathbb{U}$ agrees exactly, up to the second moment, with averaging over Haar distribution. We choose to employ the Schatten 2-norm, as it is used in the original proof of expressibility-induced kernel concentration\footnote{The final bound is phrased in terms of the Schatten 1-norm, yet the proof utilizes heavily the 2-norm and invokes Schatten norm order only in the last step to obtain results in terms of the Schatten 1-norm.} \cite{thanasilp2022exponential}. 
For notational simplicity, from now on, we fix
\begin{equation}\label{eq:expressibility}
    \varepsilon_{\mathbb{U}} = \varepsilon^{(2)}_{\mathbb{U}, 2}(\rho_0) = || \mathcal{A}^{(2)}_{\mathbb{U}}(\rho_0) ||_2.
\end{equation}
The above formula can be expressed by generalized frame potentials \cite{ali2020chaos, nakaji2021expressibility}, leading to 
\begin{equation}\label{eq:epsilon_2}
    \varepsilon^2_{\mathbb{U}} = \int_{\mathbb{U}} dU \int_\mathbb{U} dV |\langle 0 |V^{\dagger}U| 0 \rangle|^4 - \frac{1}{2^{n-1}(2^n+1)}.
\end{equation}
The integral above is calculated over the uniform volume element in the set $\mathbb{U}$ of feature maps. In our case, the set consists of unitary tranformations parameterized with the data $U(x)$. The set of transformations $\mathcal{U}_x$, for given dataset $\mathcal{X}$, can be explicitly constructed: 
\begin{equation}
\mathbb{U}_x = \{ U(x) | \forall x \in \mathcal{X}\}.
\end{equation}
In order to estimate the expressibility of the feature map for a given dataset, we follow the approach presented in \cite{holmes2022connecting}. We approximate the integrals in Eq.~\ref{eq:epsilon_2} by averaging the fidelity squared over whole $\mathbb{U}_x$:
\begin{equation}\label{eq:exp_expressibility}
    \hat{\varepsilon}_{\mathbb{U}_x} = \frac{1}{|\mathbb{U}_x|^2}\sum_{i,j}^{|\mathcal{X}|} |\langle  0 |U^{\dagger}(x_j) U(x_i) |  0 \rangle |^4 - \frac{1}{2^{n-1}(2^n+1)} \approx \varepsilon_{\mathbb{U}}.
\end{equation}

\subsection{Entanglement}
The second characteristic of the feature map describes the amount of non-local correlation between qubits. We expect that the more qubits are connected with multiple-qubit gates, the more entangled our resulting quantum system will be. Performing a partial trace on one of the subsystems of our state can be seen as ``averaging out'' this subsystem from the full state. If the initial state possessed a lot of non-local correlations (or was highly entangled), averaging out some of the degrees of freedom belonging to the subsystem would highly decrease informational content of the initial system. A reduced density matrix of a maximally entangled state is a maximally mixed state $\rho_{MMS} = \mathbb{I}/d$, where $d$ is the dimension of the reduced Hilbert space. Clearly, maximally mixed state, being an uniform superposition of states, reflects the highest possible degree of uncertainty about the state of the system and carries minimal informational content.
In quantum computing, it is believed that utilizing the resource of entanglement is a key for obtaining advantage over classical computational methods. This can pose a problem for projected quantum kernels (defined below, Eq.~\ref{eq:projected_kernel}). On the one hand, we would like to use the resource of entanglement. On the other hand, when our functions are based on reducing a highly entangled system, then the reduced density matrices that we obtain will be really close to maximally mixed state $\rho_{MMS}$.
Having that in mind, in this paper, we will measure the amount of entanglement utilizing quantum relative entropy with respect to maximally mixed state
\begin{equation}\label{eq:quantum_relative_entropy}
    S(\rho_k || \rho_{MMS}) = \Tr[\rho_k(\log \rho_k - \log \rho_{MMS})],
\end{equation}
where $\rho_k$ is the reduced density matrix.
A good indication that the embedded state (Eq.~\ref{eq:embedded_state}) was a highly entangled state would be if $S(\rho_k || \rho_{MMS})$ is close to zero.

\subsection{Characteristics of the ZZ-Feature Map}
\begin{figure}[ht!]
    \centering
    \includegraphics[width = \textwidth]{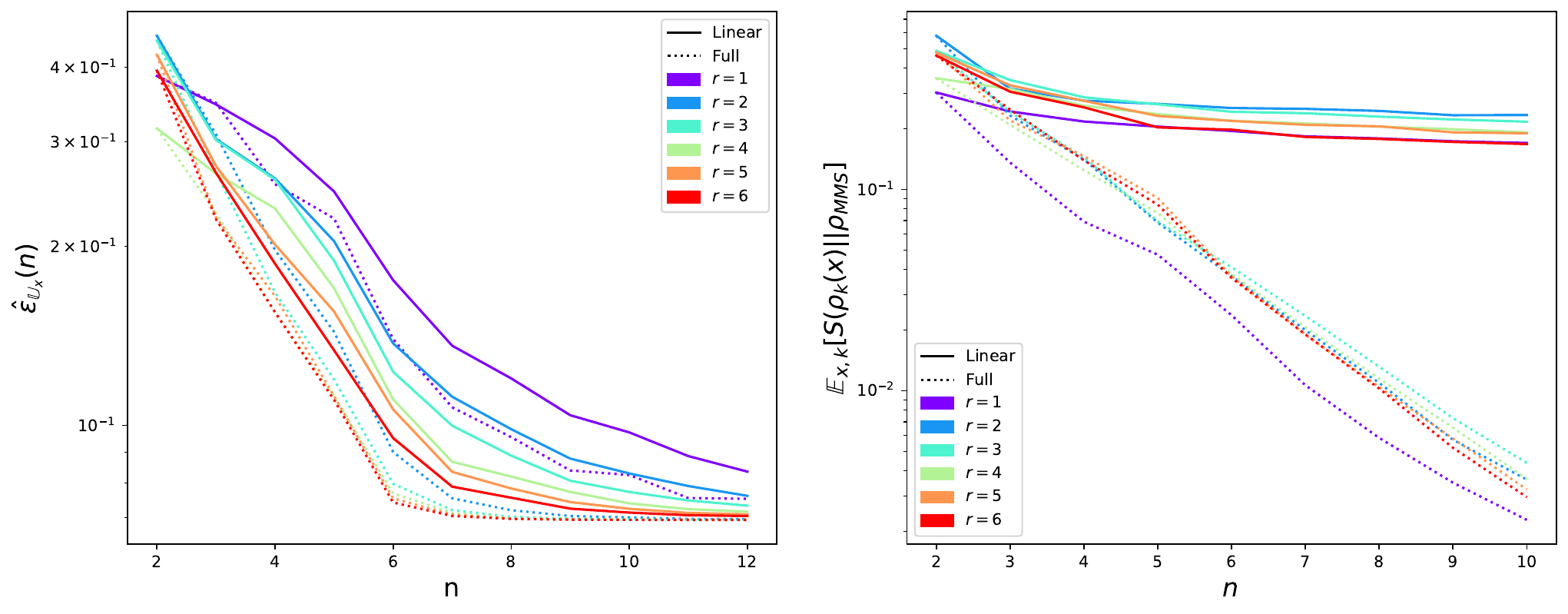} 
    \caption{The two characteristics of the ZZ-feature map, the numerically obtained measure of expressibility (Eq.~\ref{eq:exp_expressibility}) and the measure of entanglement in the terms of relative entropy (Eq.~\ref{eq:quantum_relative_entropy}). \textit{Left:} The expressibility measure of a $r$-times repeated ZZ-feature map as a function of number of qubits $n$ for both fidelity (Eq.~\ref{eq:fidelity_kernel}) and projected (Eq.~\ref{eq:projected_kernel}) quantum kernels. The measure does not vanish exponentially, but rather tends to a constant value of about $0.07$. In general, the projected quantum kernel expressibilities are similar to the fidelity quantum kernel ones. \textit{Right: } The mean quantum relative entropy between reduced density matrices and maximally mixed state density matrix for projected quantum kernels as a function of number of qubits $n$. Observe that, the linear entanglement scheme in feature map does not lead to the exponential vanishing of the said measure.}
            \label{fig:expr_ent}
\end{figure}
In order to characterize the behavior of the used feature map, we numerically simulated the expressibility (Eq.~\ref{eq:expressibility}) and entanglement (Eq.~\ref{eq:quantum_relative_entropy}) measures as a function of number of qubits $n$ and number of repetitions $r$.
The results are shown in Figure~\ref{fig:expr_ent}---all plots shown in this chapter are obtained for the \textit{Connectionist Bench} dataset (the data points were normalized)~\cite{sonar}. 
We exploit a simple feature selection method in the data preprocessing step---the features were ordered with decreasing variance. 
The size of the quantum machine was increased by selecting successive features, starting with two first (greatest variance) and finishing with first 12 features (out of $60$ all features).
In the simulations, we have used multiple repetitions $r\in \{1,2,3,4,5,6\}$ of the ZZ-feature map with the \textit{linear} and \textit{full} entanglement strategy.

The first observation is that with a growing number of qubits $n$ and number of repetitions $r$ of the feature map, the expressibility increases, as expected.
For $r>3$, the expressibility measure (Eq.~\ref{eq:exp_expressibility}) at first does vanish exponentially, but around $n=7$ hits a plateau. 
We conjecture that this is not an inherent property of the feature map used, but the limited number of samples in the dataset used ($m=208$). The feature map is not able to explore reliably a huge feature space.

For the entanglement, the right plot of Figure~\ref{fig:expr_ent} shows a mean quantum relative entropy between a one-qubit reduced density matrix and a maximally mixed state $\mathbb{I}/2$. The mean is calculated, first over all possible 1-qubit reduced states of the embedded data, and then over all Connectionist Bench data points. 
One can see a clear distinction between the linear and full entanglement schemes. Full entanglement exhibits exponential decay with an increasing qubit number $n$. Linear entanglement scheme shows a different behavior, it decreases slightly for small number of qubits and hits a plateau quite quickly (around $n=5$ qubits) on which the quantum relative entropy value is approximately constant.
An interesting observation is that increasing the number of repetitions $r$ does not change the behavior for both of the entanglement schemes.
Therefore, we can draw a conclusion of separation of effects. Expressibility is mostly affected by number of feature map repetitions $r$, while entanglement is affected by the entanglement scheme, as expected.





\section{Number of Shots Estimation for Quantum Fidelity and Projected Quantum Kernels}\label{App:N_estimates}

\subsection{Spread Comparison}
This bound is based on the assumption that we would like to have a lower uncertainty of the kernel value estimation than the characteristic spread in the kernel values in the ensemble (Eq.~\ref{eq:precision_ratio} in the main body of the paper):
\begin{equation}
      \varepsilon = \frac{\Delta_{kernel\ value}}{\Delta_{ensemble}}  \lesssim 1 .
\end{equation}
Identifying $\Delta_{kernel\ value}= |\hat{\kappa} - \kappa|$, where $\kappa = \mathbb{E}[\hat{\kappa}]$ we can estimate the number of shots $N$ from the probability of measuring a kernel value that is $\varepsilon \Delta_{ensemble}$ from its true value
\begin{equation}
    Pr[|\hat{\kappa} - \kappa| \geq \epsilon \Delta].
\end{equation}
\subsubsection{Fidelity Quantum Kernel}
Using Chebyshev's inequality we have
\begin{equation}\label{eq:Fidelity_Chebyshev}
    Pr[|\hat{\kappa} - \kappa| \geq \epsilon \Delta_{ensemble}] \leq \frac{Var[\hat{\kappa}]}{(\varepsilon \Delta_{ensemble})^2}.
\end{equation}
Now, remember that for the fidelity quantum kernel the kernel value is directly a binomial proportion having the well known variance 
\begin{equation}
Var[\hat{\kappa}] = \frac{1}{N} \kappa(1-\kappa).
\end{equation}
Using the transitivity property on the inequality (\ref{eq:Fidelity_Chebyshev}) we can introduce a probability $(1-P)$ which can serve as an upper bound of the rigth-hand side of (\ref{eq:Fidelity_Chebyshev}). We then obtain
\begin{equation}\label{eq:spread_FQ_bound}
    N_{spread} \geq \frac{\kappa(1-\kappa)}{(1-P)(\varepsilon \Delta_{ensemble})^2}.
\end{equation}

\subsubsection{Projected Quantum Kernel}

First let us introduce notation for basic random variables
\begin{equation}
\hat{Z}^{(k)} \in \{\hat{\rho}_{00,k}(x), \Re(\hat{\rho}_{01,k})(x), \Im(\hat{\rho}_{01,k})(x), \hat{\rho}_{00,k}(y), \Re(\hat{\rho}_{01,k})(y), \Im(\hat{\rho}_{01,k})(y) \},
\end{equation}
in such a way that $\hat{Z}_i^{(k)}, i\in \{1,\dots, 6\}$, and, for example, $\hat{Z}_5^{(k)} = \Re(\hat{\rho}_{01,k})(y)$.
Moreover, the random variables are written with hats, while their expectation values are without hats, $\mathbb{E}[\hat{Z}_i^{(k)}] = Z_i^{(k)}$.
Remember, that each of the basic variables comes from the $N$-trial Bernoulli process, hence we know that
\begin{equation}
    Var[\hat{Z_i^{(k)}}] = \frac{1}{N} Z_i^{(k)}\cdot(1-Z_i^{(k)}).
\end{equation}

\noindent The projected quantum kernel used in the main document is defined as
\begin{equation}
    \hat{\kappa} = e^{-\gamma \sum_{k=1}^n || \hat{\rho}_k(x) - \hat{\rho}_k(y) ||^2_2}.
\end{equation}
The Schatten 2-norm of the difference between reduced density matrices is
\begin{align}\label{eq:norm_reduced}
    || \hat{\rho}_k(x) - \hat{\rho}_k(y) ||^2_2 &= 2 \{ [\hat{\rho}_{00,k}(x)-\hat{\rho}_{00,k}(y)]^2 + [\Re(\hat{\rho}_{01,k}(x))-\Re(\hat{\rho}_{01,k}(y))]^2 + [\Im(\hat{\rho}_{01,k}(x))-\Im(\hat{\rho}_{01,k}(y))]^2 \}\\
    & = 2 \{ [\hat{Z}_{1}^{(k)}-\hat{Z}_{4}^{(k)}]^2 + [\hat{Z}_{2}^{(k)}-\hat{Z}_{5}^{(k)}]^2 + [\hat{Z}_{3}^{(k)}-\hat{Z}_{6}^{(k)}]^2 \}
\end{align}
In order to perform a step-by-step analysis, we introduce 
\begin{align}
    \hat{X} &= \sum_{k=1}^n || \hat{\rho}_k(x) - \hat{\rho}_k(y) ||^2_2 = \sum_{k=1}^n \hat{X}^{(k)}
\end{align}
Similarly as above, we would like to use a Chebyshev inequality in order to restrict the probability that the measured projected kernel entries are $\Delta$ from their expected value $\mathbb{E}[\hat{\kappa}] = \kappa$, additionally
\begin{equation}\label{eq:chebyshev}
    Pr[|\hat{\kappa} - \kappa| \geq \Delta] \leq \frac{Var[\hat{\kappa}]}{\varepsilon\Delta_{ensemble}^2} \approx \left( \hat{\kappa}'[X] \right)^2 \frac{Var[\hat{X}]}{\varepsilon\Delta_{ensemble}} \leq ...,
\end{equation}
where the approximation comes from the use of the delta method.
The variables $X^{(k)}$ come from partially tracing out the same matrix, therefore we treat them as possibly correlated variables. We know that
\begin{equation}\label{eq:varx}
    Var[\hat{X}] = Var \left[ \sum_k \hat{X}^{(k)} \right] \leq n \sum_k Var[\hat{X}^{(k)}].
\end{equation}
Now, we use again the delta method, in this case multivariate
\begin{equation}
    Var[\hat{X}^{(k)}] \approx \sum_i \left( \frac{\partial \hat{X}^{(k)}}{\partial \hat{Z}_i^{(k)}} \right)^2 Var[\hat{Z}_i^{(k)}] + \sum_{i\neq j} \left( \frac{\partial \hat{X}^{(k)}}{\partial \hat{Z}_i^{(k)}} \right)\left( \frac{\partial \hat{X}^{(k)}}{\partial \hat{Z}_j^{(k)}} \right)Cov[\hat{Z}_i^{(k)},\hat{Z}_j^{(k)}].
\end{equation}
Using triangle inequality we get
\begin{align}\label{eq:triange_inequality}
    Var[\hat{X}^{(k)}] = |Var[\hat{X}^{(k)}]| \lessapprox \sum_i \left( \frac{\partial \hat{X}^{(k)}}{\partial \hat{Z}_i^{(k)}} \right)^2 Var[\hat{Z}_i^{(k)}] + \sum_{i\neq j} \bigg\rvert\left( \frac{\partial \hat{X}^{(k)}}{\partial \hat{Z}_i^{(k)}} \right) \left( \frac{\partial \hat{X}^{(k)}}{\partial \hat{Z}_j^{(k)}} \right)\bigg\rvert |Cov[\hat{Z}_i^{(k)},\hat{Z}_j^{(k)}]|.
\end{align}
As was mentioned in the begining of this appendix, we know variances of basic random variables. For covariances one can use Cauchy-Schwarz inequality
\begin{equation}
    |Cov[\hat{Z}_i^{(k)},\hat{Z}_j^{(k)}]| \leq \sqrt{Var[\hat{Z}_i^{(k)}] Var[\hat{Z}_j^{(k)}]} = \frac{1}{N}\sqrt{\hat{Z}_i^{(k)}(1-\hat{Z}_i^{(k)})\hat{Z}_j^{(k)}(1-\hat{Z}_i^{(k)})}.
\end{equation}

\noindent Coming back to the equation (\ref{eq:chebyshev}) we use (\ref{eq:varx}) and (\ref{eq:triange_inequality}) to get
\begin{align}
    ... &\leq \left( \hat{\kappa}'[X] \right)^2 \frac{n}{(\varepsilon\Delta_{ensemble})^2} \sum_k Var[X^{(k)}] \leq \\
    &\leq \left( \hat{\kappa}'[X] \right)^2 \frac{n}{N (\varepsilon\Delta_{ensemble})^2}\sum_k V_k,
\end{align}
where,
\begin{equation}
    V_k = \sum_i \left( \frac{\partial \hat{X}^{(k)}}{\partial \hat{Z}_i^{(k)}} \right)^2 \hat{Z}_i^{(k)}(1-\hat{Z}_i^{(k)}) + \sum_{i\neq j} \bigg\rvert\left( \frac{\partial \hat{X}^{(k)}}{\partial \hat{Z}_i^{(k)}} \right) \left( \frac{\partial \hat{X}^{(k)}}{\partial \hat{Z}_j^{(k)}} \right)\bigg\rvert | \sqrt{\hat{Z}_i^{(k)}(1-\hat{Z}_i^{(k)})\hat{Z}_j^{(k)}(1-\hat{Z}_i^{(k)})} |.
\end{equation}
Now, if we restrict the above equation with the small probability $(1-P)$, we have
\begin{equation}\label{eq:spread_PQ_bound}
    N \geq \frac{n \gamma^2\kappa^2 }{(\varepsilon\Delta_{ensemble})^2 (1-P_{target})}\sum_k V_k
\end{equation}

In order to obtain the above equation with the explicit form of $V$, we use the equation (\ref{eq:norm_reduced}) and observe that
\begin{align}
    \bigg\rvert\left( \frac{\partial \hat{X}^{(k)}}{\partial \hat{Z}_i^{(k)}} \right)\bigg\rvert &= 4|Z_i^{(k)}-Z_{i+3}^{(k)}|,\ for\ i=1,2,3\\
    \bigg\rvert\left( \frac{\partial \hat{X}^{(k)}}{\partial \hat{Z}_i^{(k)}} \right)\bigg\rvert &= 4|Z_{i-3}^{(k)}-Z_{i}^{(k)}|,\ for\ i=4,5,6.\\
\end{align}
Therefore we have
\begin{align}
    \frac{V_k}{16} &= \sum_{i=1}^3\sum_{j=1}^3 |(Z_i^{(k)}-Z_{i+3}^{(k)})(Z_j^{(k)}-Z_{j+3}^{(k)})| \sqrt{Z_i^{(k)}(1-Z_i^{(k)})Z_j^{(k)}(1-Z_j^{(k)})} \\
    &+ \sum_{i=4}^6\sum_{j=4}^6 |(Z_{i-3}^{(k)}-Z_{i}^{(k)})(Z_{j-3}^{(k)}-Z_{j}^{(k)})| \sqrt{Z_i^{(k)}(1-Z_i^{(k)})Z_j^{(k)}(1-Z_j^{(k)})} \\
    &+ 2\sum_{i=1}^3\sum_{j=4}^6 |(Z_i^{(k)}-Z_{i+3}^{(k)})(Z_{j-3}^{(k)}-Z_{j}^{(k)})| \sqrt{Z_i^{(k)}(1-Z_i^{(k)})Z_j^{(k)}(1-Z_j^{(k)})}.
\end{align}


\subsection{Concentration Avoidance}
Knowing that the quantum measurement outcomes will concentrate around specific value, we wish to make sure that for the expected measurement value $\mathcal{M}$ which is different from the concentrated measurement value $\mu_{\mathcal{M}}$ we have a $P_{CA}$ chance that the measured value $\hat{\mathcal{M}}$ will be different from $\mu_{\mathcal{M}}$.
The derivation of the bound will mathematically be identical, but as the issue addresses the measurement, rather than directly kernel value, the implementation of the bound will differ for FQK and PQK kernel families.

This idea connects directly to the measurement process in the quantum kernel estimation procedure, therefore will differ for fidelity-based kernels and for kernels for which we perform quantum one-qubit tomography.

Additionally, if $\mathcal{M} = \mu_{\mathcal{M}}$ we do not impose the bound.

Colloquially, we would like to \textit{push away} the measurement from the concentrated value. 
In mathematical terms,
\begin{itemize}
    \item if $\mathcal{M} < \mu_{\mathcal{M}}$\\ 
    we would like to measure $\hat{\mathcal{M}} < \mu_{\mathcal{M}}$ with probability $P_{CA}$,
\begin{equation}
Pr[\hat{\mathcal{M}} < \mu_{\mathcal{M}}] \geq P_{CA}.
\end{equation}
Remembering that the measurement boils down to the estimation of proportion in binomial distribution in $N$ trials, we can rewrite the above equation as
\begin{equation}
    Pr[k < N\cdot \mu_{\mathcal{M}}, N, \mathcal{M}] \geq P_{CA},
\end{equation}
which then can be rephrased in the terms of Cumulative Distribution Function (CDF),
\begin{equation}
    CDF(k = (N \mu_{\mathcal{M}}-1)) \geq P_{CA},
\end{equation}
where in our case CDF can be identified with the regularized incomplete beta function $CDF(k) = I_{1-\mathcal{M}}(N-\lfloor k \rfloor, 1+\lfloor k \rfloor)$.
    \item if $\mathcal{M} > \mu_{\mathcal{M}}$\\
    we have the bound in the following form
    \begin{equation}
        Pr[\hat{\mathcal{M}} > \mu_{\mathcal{M}}] \geq P_{CA}.
    \end{equation}
    Again, the bound can be rephrased with the help of the CDF
    \begin{equation}
        1-CDF(k = N \mu_{\mathcal{M}}) \geq P_{CA}
    \end{equation}
\end{itemize}
Collecting the above results we have
\begin{equation}
    \begin{cases}\label{eq:sr_bound}
        CDF(k = (N_{SR} \mu_{\mathcal{M}}-1)) \geq P_{CA}, & \text{for } \mathcal{M} < \mu_{\mathcal{M}}\\
        1-CDF(k = N_{SR} \mu_{\mathcal{M}}) \geq P_{CA}, & \text{for } \mathcal{M} > \mu_{\mathcal{M}}
    \end{cases}
\end{equation}
Below we discuss how to apply the above bounds to the specific cases of kernel families.

\subsubsection{Fidelity Quantum Kernel}
In the case of the fidelity quantum kernel the measured value is exactly the kernel value and is equal to
\begin{equation}
    \mu_{\mathcal{M}} = 0.
\end{equation}
From the definition of the kernel we know that $\mathcal{M} = Tr[\rho(x)\rho(y)] \geq 0$, hence in the estimation of $N$ we will consider only the case $\mathcal{M} > \mu_{\mathcal{M}}$. 
Moreover, for this specific case of $\mathcal{M}$ and $\mu_{\mathcal{M}}$ we can substantially simplify the cumulative distribution function
\begin{equation}
    CDF(k=0) = I_{1-\mathcal{M}}(N,1) = Bin(k=0,N,\mathcal{M}) = (1-\mathcal{M})^N.
\end{equation}
That can be then applied to the bound (\ref{eq:sr_bound}), giving the simplified formula
\begin{equation}\label{eq:sr_fidelity_bound}
    N \geq log_{1-\mathcal{M}}(1-P_{CA}),
\end{equation}
which then can be applied to every kernel matrix value. Identifying $\mathcal{M} = \kappa^{FQ}$ and expanding the left-hand side of the above equation with respect to small $\kappa^{FQ}$ gives the intuitive behavior $N^{FQ} \sim ln(\frac{1}{1-P_{CA}}) [\kappa^{FQ}]^{-1}$.
Here, we make a comment about calling the investigated issue a \textit{success rate}. This interpretation stems from the fidelity quantum kernel, for which, if we do not have high enough $N$, we might run into the issue of measuring only \textit{failures} in all the trials from binomial distribution. Hence, the bound (\ref{eq:sr_fidelity_bound}) enforces the probability $P_{CA}$ of obtaining at least one \textit{success} in the measurement process.

\subsubsection{Projected Quantum Kernel}

In the case of the projected quantum kernel, we are measuring independent entries of the reduced $1$-qubit density matrices which then can be translated to $\mathcal{M}$'s (compare with (\ref{eq:projected_rhos})
\begin{align}
    \begin{split}
        \mathcal{M}_k^D &= \rho^D_k, \\
        \mathcal{M}_k^R &= \rho^R_k + \frac{1}{2}, \\
        \mathcal{M}_k^I &= \frac{1}{2} - \rho^I_k.
    \end{split}
\end{align}
The measured values $\mathcal{M}$'s are all concentrated around
\begin{equation}
    \mu = \frac{1}{2}.
\end{equation}
In this case the CDF does not simplify substantially and one has to solve the bounds (\ref{eq:sr_bound}) by numerically inverting the regularized incomplete beta function.

Additionally, observe that the measurement for projected quantum kernels is not performed for a kernel value, but for the specific datapoint. The kernel values are then obtained in the classical post-processing. For each data point we will have $3 \cdot n$ estimates of the $N$ value, where $n$ is the number of logical qubits of the quantum device.

\section{Noisy Circuits}\label{App:noisy_circuits_new}
In this appendix we estimate the acceptable level of error in implementing the quantum circuit for the fidelity and projected quantum kernels. 

\subsection{Our Noise Model: Depolarizing}\label{app:noise_model}
We first introduce the noise model we consider in our estimates. We call $\rho$ the final density matrix at the end of the quantum circuit in a noiseless scenario. Because of noise, we cannot have exactly access to $\rho$: instead we have access to a density matrix $\rho_f$. We assume that with a probability $p$, an error occured in the circuit and gave a wrong density matrix $\widetilde{\rho}$ instead of $\rho$. Hence, the final density matrix in the circuit, $\rho_f$, reads:
\begin{align}
    \rho_f=(1-p) \rho+p \widetilde{\rho}.
    \label{eq:rho_f}
\end{align}
For our quantitative results, we will further assume that $\widetilde{\rho}=\mathbb{I}/2^n$, hence that the noise model is depolarizing.

\subsection{Fidelity Quantum Kernel}
\label{app:def_problem}
\subsubsection{Problem Definition}
We recall that, for the fidelity Quantum Kernel, the goal is to estimate $\Tr \left[|0\rangle \langle 0 | \rho \right]$. In a noiseless case, this quantity is simply estimated by counting the number of times the final measurement gives the outcome $|0\rangle$. In a noisy case, we will also estimate this quantity by counting the number of times the final state collapses to $|0\rangle$, but as the final state is $\rho_f$ and not $\rho$, we will introduce an error in our estimate. To simplify notations in the rest of the text, we introduce:
\begin{align}
    \kappa \equiv \Tr \left[|0\rangle \langle 0 | \rho \right]
\end{align}

Even if the circuit was noiseless, we could not \textit{exactly} access $\kappa$.
Rather, we would have an estimator of this quantity, which is defined as:
\begin{align}
    \hat{\kappa} = \sum_{i=1}^{N} \frac{X_i}{N},
\end{align}
where $X_i$ is a random variable which returns $1$ if the measurement collapses $\rho$ to $| 0 \rangle$ and $0$ otherwise.
The variables $\{X_i\}_{i=1}^{N}$ are statistically independent (and follow the same distribution probability) because circuit runs are uncorrelated. The quantity $N$ is the number of circuit repetitions. 

When the circuit is noisy, we have an estimator $\hat{\kappa}_f$ which is defined in a very similar manner:
\begin{align}
    \hat{\kappa}_f = \sum_{i=1}^{N} \frac{Y_i}{N},
\end{align}
Here, $Y_i$ is also a random variable returning $1$ if $\rho_f$ has collapsed to $|0\rangle$, and $0$ otherwise. The variables $\{Y_i\}_{i=1}^{N}$ are also statistically independent (and follow the same distribution probability) because we assume the circuit runs are uncorrelated. 
The difference between $\hat{\kappa}$ and $\hat{\kappa}_f$ relies in the fact that with $\hat{\kappa}_f$ the circuit is noisy, which means that the probability followed by $Y_i$ is different from the one followed by $X_i$ (this is because the final density matrix is $\rho$ in the noiseless case, and $\rho_f$ in the noisy one: the measurement outcome distribution is then different).
Again, observe that,
\begin{align}
    \mathbb{E}[\hat{\kappa}_f] = \Tr[|0\rangle \langle 0| \rho_f].
\end{align}

Our goal here is to guarantee that for a chosen $\Delta>0$, we have:
\begin{align}\label{eq:delta_estimator}
    |\hat{\kappa}_f-\kappa| \leq \Delta,
\end{align}
with a probability being at least equal to $P_{\text{target}}$. 

Hence, we want to find an upper bound on $p$ that will guarantee us that \eqref{eq:delta_estimator} is true (for some $\Delta$ of our choice), with a probability being at least $P_{\text{target}}$ (for a $P_{\text{target}}$ of our choice). It is this upper bound that will allow us to estimate error-correction overheads (as we would know how often we can allow the circuit to fail, which is quantified by $p$). In practice, a good value of $\Delta$ and $P_{\text{target}}$ depends on the precision we want for the task. 

\subsubsection{Finding an Upper Bound on $p$}\label{app:subsection_p}

Our goal now is to find an upper bound on $p$ in \eqref{eq:rho_f} such that \eqref{eq:delta_estimator} is satisfied with a probability greater than $P_{\text{target}}$. Using the fact that $\mathbb{E}[\hat{\kappa}]=\kappa$, we can rewrite \eqref{eq:delta_estimator} as
\begin{align}
    &|\hat{\kappa}_f-\kappa| = |\hat{\kappa}_f-\mathbb{E}[\hat{\kappa}]|= |(\hat{\kappa}_f - \mathbb{E}[\hat{\kappa}])-\Delta \mu + \Delta \mu| \label{eq:DeltaNcN0},\\
    &\Delta \mu \equiv \mathbb{E}[\hat{\kappa}_f]-\mathbb{E}[\hat{\kappa}].
    \label{eq:DeltaNcN1}
\end{align}

The quantity $\Delta \mu$ represents the difference in average value of the estimator $\hat{\kappa}_f$ and $\hat{\kappa}$. Now, defining the random variable $Z = \hat{\kappa}_f - \mathbb{E}[\hat{\kappa}]$, we have $\mathbb{E}[Z]=\Delta \mu$. Hence, from \eqref{eq:DeltaNcN0}, we have
\begin{align}
    |\hat{\kappa}_f-\kappa| =|Z-\mathbb{E}[Z]+\Delta \mu| \leq |Z-\mathbb{E}[Z]|+|\Delta \mu|.
\end{align}
It indicates that if we can guarantee the following two inequalities, then \eqref{eq:delta_estimator} is satisfied
\begin{align}
    &|\Delta \mu| \leq \frac{\Delta}{2} \label{eq:cond:deltamu},\\
    &|Z-\mathbb{E}[Z]| \leq \frac{\Delta}{2} \label{eq:cond:Z}.
\end{align}
Additionally, we have
\begin{align}
    |\Delta \mu|&=|\Tr \left[| 0 \rangle\langle 0 | \rho_f \right]  - \kappa |, \\
    &=p | \Tr \left[| 0 \rangle\langle 0 | \widetilde{\rho} \right] - \kappa |.
\end{align}
It gives us an inequality allowing to satisfy the first of the two conditions, \eqref{eq:cond:deltamu}
\begin{align}
    |\Delta \mu| \leq \frac{\Delta}{2} \Leftrightarrow p \leq \frac{\Delta}{2| \Tr \left[| 0 \rangle\langle 0 | \widetilde{\rho} \right]-\kappa |}.
\end{align}
It remains for us to satisfy \eqref{eq:cond:Z} with a probability being at least $P_{\text{target}}$. From Chebyshev's inequality, we have
\begin{align}
    P(|Z-\mathbb{E}[Z]| \geq \frac{\Delta}{2}) \leq \frac{Var(Z)}{(\frac{\Delta}{2})^2}.
    \label{eq:Chebyshev}
\end{align}
We also know that
\begin{align}
    Var(Z)=Var(\hat{\kappa}_f)=\frac{1}{N^2}\sum_i Var(Y_i)=\frac{1}{N} Var(Y_i) \leq \frac{1}{N}.
    \label{eq:varZ}
\end{align}
There, we used the facts (i) the $\{Y_i\}$ are independent variables following the same probability distribution, (ii) $Var(Y_i) \leq 1$ (because $Y_i \in [0,1]$).
Contrary to the Appendix \ref{App:N_estimates} in the case of noisy estimator $\hat{\kappa}_f$ we cannot use the analytic formula $Var[\hat{\kappa}] = \frac{\kappa(1-\kappa)}{N}$. This leads to much less strong bound on the variance of our random variable.

In order to satisfy \eqref{eq:cond:Z} with a probability at least $P_{\text{target}}$, it is sufficient to have
\begin{align}
    1-\frac{1}{N (\frac{\Delta}{2})^2} \geq P_{\text{target}}.
\end{align}
To obtain it we simply combined \eqref{eq:Chebyshev} and \eqref{eq:varZ}.

Our goal is to guarantee that for some error $\Delta \geq 0$, we have
\begin{align}
    |\hat{\kappa}_f-\kappa| \leq \Delta
\end{align}
with a probability being at least $P_{\text{target}}$, and from that deduce a maximum acceptable rate for the circuit, $p$. We showed that this condition is satisfied if we can guarantee that the following two inequalities are true, with a probability being at least $P_{\text{target}}$
\begin{align}
    &|\Delta \mu| \leq \frac{\Delta}{2}, \\
    &|Z-\mathbb{E}[Z]| \leq \frac{\Delta}{2}.
\end{align}
Such inequalities are satisfied if,
\begin{align}
&p \leq \frac{\Delta}{2| \Tr \left[| 0 \rangle\langle 0 | \widetilde{\rho} \right]-\kappa |} \label{eq:final_cond_1},\\
&1-\frac{1}{N (\frac{\Delta}{2})^2} \geq P_{\text{target}} \label{eq:final_cond_2}.
\end{align}
In this way we obtain two conditions. 

The condition (\ref{eq:final_cond_1}) governs the probability $p$ that an error occurred in the execution of the algorithm. We will treat the upper bound on $p$ as an error budget for one run of the algorithm for the error-correction resource estimate (see section \ref{App:error_estimates}).  In practice, the right-hand side of \eqref{eq:final_cond_1} requires us to know $\Tr \left[| 0 \rangle\langle 0 | \widetilde{\rho} \right]$ and $\kappa$. 
Here, we use the fact that the global effect of the noise on the circuit is depolarizing (see section \ref{app:noise_model}), meaning that $\widetilde{\rho}=\mathbb{I}/2^n$. For this reason, $\Tr \left[| 0 \rangle\langle 0 | \widetilde{\rho} \right] = 1/2^n$. Following the spirit of the main document, for resource estimation we use (\ref{eq:precision_ratio}), (\ref{eq:sr}) and take $\Delta = \varepsilon \Delta_{ensemble} = \varepsilon IQR(\mathbb{K})$, $\kappa=median(\mathbb{K})$. When performing quantum resource estimation beyond our simulation capabilities, we perform an exponential extrapolation of values of $\Delta$ and $\kappa$. The values are $log_2$ transformed, and fitted with the standard linear regression. If the fit meets the quality threshold, $R^2=0.99$, then we use the obtained $C e^{\alpha \cdot n}$ function for extrapolation.

The condition (\ref{eq:final_cond_2}) plays no role in the error-correction overheads: it will only allow us to estimate the number of times the circuit will be repeated, hence its duration and total energy cost.

\subsection{Projected Quantum Kernels}

The condition \eqref{eq:final_cond_2} for projected quantum kernels can be obtained by utilizing the results from Appendix \ref{App:N_estimates}. Using the same line of thoughts as above, definition of projected quantum kernel and the delta method one obtains
\begin{equation}
1-\frac{n \gamma^2\kappa^2 }{N(\frac{\Delta}{2})^2}\sum_k \hat{V}_k \geq P_{target},
\end{equation}
with modified
\begin{equation}
    \hat{V}_k = \sum_{i, j} \bigg\rvert\left( \frac{\partial \hat{X}^{(k)}}{\partial \hat{Z}_i^{(k)}} \right) \left( \frac{\partial \hat{X}^{(k)}}{\partial \hat{Z}_j^{(k)}} \right)\bigg\rvert,
\end{equation}
where, again, we had to use the variance bound (\ref{eq:varZ}).

In the case of the condition similar to (\ref{eq:final_cond_1}) we will follow the reasoning presented below.
First let's focus on the expected kernel entry obtained from the noisy circuit
\begin{equation}
    \kappa_f = e^{-\gamma \sum_k ||\rho_{f,k}(x)-\rho_{f,k}(y)||^2_2}.
\end{equation}
For convenience we introduce the following notation $\Delta\rho_{f,k} = \rho_{f,k}(x)-\rho_{f,k}(y) = (1-p)\Delta \rho_k + p \Delta \Tilde{\rho}_k$, where $\Delta\rho_k = \rho_k(x)-\rho_k(y)$ and $\Delta\Tilde{\rho}_k = \Tilde{\rho}_k(x)-\Tilde{\rho}_k(y)$.
The 2-norm of the different of density matrices can be rewritten as
\begin{equation}
    || \Delta \rho_{f,k}||^2_2 = (1-p)^2 || \Delta \rho_k ||^2_2 + p^2 || \Delta \Tilde{\rho}_k||^2_2 + 2p(1-p)\langle \Delta \rho_k, \Delta \Tilde{\rho}_k \rangle_F,
\end{equation}
where $\langle \cdot, \cdot \rangle_F$ is a Frobenius product.
Now, we will perform approximations, first of all, we expect that the allowed probability of error is small $p \ll 1$ and we will drop $\mathcal{O}(p^2)$ terms 
\begin{align}
    \kappa_f = \exp{-\gamma \sum_k ||\Delta \rho_{f,k}||^2_2} &\approx \exp{-\gamma (1-2p) \sum_k ||\Delta \rho_{k}||^2_2-2\gamma p \sum_k \langle \Delta\rho_k, \Delta\Tilde{\rho}_k \rangle_F} \\
    &\approx \underbrace{\exp{-\gamma \sum_k ||\Delta \rho_{k}||^2_2}}_{\kappa} \cdot \left(1-2\gamma p \sum_k\left( Re[\langle \Delta\rho_k, \Delta\Tilde{\rho}_k \rangle_F] - ||\Delta\rho_k||^2_2 \right)  \right)
\end{align}
where in the second line we have expanded the exponential functions to the first order in $p$.
For the condition for the error budget we have
\begin{equation}
|\kappa_f - \kappa| = 2p\gamma\kappa\bigr\rvert \sum_k(||\Delta \rho_k||^2_2)-Re[\langle \Delta\rho_k, \Delta\Tilde{\rho}_k \rangle_F] \bigr\rvert \leq \frac{\Delta}{2}
\end{equation}
\begin{equation}
    p \lessapprox \frac{\Delta}{4\gamma\kappa \bigr\rvert \sum_k(||\Delta \rho_k||^2_2)-Re[\langle \Delta\rho_k, \Delta\Tilde{\rho}_k \rangle_F] \bigr\rvert}.
\end{equation}
In the depolarizing noise model $\mathbb{E}\left[\Delta \Tilde{\rho}_k\right] = 0$, therefore in the linear approximation we can simplify further
\begin{equation}
    p \lessapprox -\frac{\Delta}{4\kappa \ln\kappa}.
    \label{eq:p_projected}
\end{equation}

\noindent Again, as in Section \ref{app:subsection_p}, we identify the variables on the right-hand side of the above equation to be $\Delta = \varepsilon \Delta_{ensemble} = \varepsilon IQR(\mathbb{K})$ and $\kappa=median(\mathbb{K})$, and we resort to extrapolation for the values beyond our computational range.
\subsection{Concentration Avoidance Bound for the Noisy Devices}
In the concentration avoidance bound, we are interested in ensuring that the number of circuit runs we perform is sufficient to distinguish the estimated kernel entry from value to which it concentrates. 
In the case of noisy devices, there is a finite probability that the error occurs in the execution of the circuit.
In every circuit run, the estimated kernel value $\hat{\kappa}_f$ might be different.
Therefore, instead of considering the binomial distribution, we have to go a step back and consider a collection of $N$ measurements according to Bernoulli distribution with probability of ``success'' treated as a random variable $\{\hat{\mathcal{M}}_i\}_{i=1}^{N}$.
For a random variable $\hat{X}_i \in \{0,1\}$ which is the result of $i^{th}$ measurement, and random variable $\hat{k} = \hat{X}_1 + \dots + \hat{X}_{N}$ we would like to bound the probability of obtaining more ``successes'' in the consecutive Bernoulli trials than some threshold $C$ based on the concentration value
\begin{align}\label{app:noisy_sr_bound}
    Pr[\hat{k}>C] \geq P_{CA}.
\end{align}
We know that,
\begin{align}
    Pr[\hat{k}>C] = Pr\left[\bigcup\limits_{l=C+1}^{N} \hat{X}_1+\dots+\hat{X}_{N} = l\right] = \sum_{l=C+1}^{N} Pr[\hat{X}_1+\dots+\hat{X}_{N} = l],
\end{align}
where in the last equality we have used the fact that the collection of measurement results summing to specific value are mutually exclusive.
We introduce the set $\Sigma_l(N)$ of all combinations of values of $\hat{X}_1,\dots,\hat{X}_{N}$, that satisfy $\hat{X}_1+\dots+\hat{X}_{N} = l$. The set consists of $|\Sigma_l(N)| = \frac{N!}{l!(N-l)!}$ elements. With the set element $\sigma \in \Sigma_l(N)$ we introduce index notation $\sigma_i$ where first $l$ indices indicate random variables $\hat{X}_{\sigma_i}$'s which were measured as a `success', while the latter $N-l$ indices indicate variables which were measured as a `failure', $\hat{X}_{\sigma_1}+ \dots + \hat{X}_{\sigma_l} = l$ and $\hat{X}_{\sigma_{l+1}}+ \dots + \hat{X}_{\sigma_{N}} = 0$.
Now, for mutually exclusive events we have
\begin{align}
    Pr[\hat{X}_1+\dots+\hat{X}_{N} = l] = \sum_{\sigma \in \Sigma_l(N)} Pr[\sigma],
\end{align}
and
\begin{align}
    Pr[\sigma] = \prod_{i=1}^l \prod_{j=l+1}^{N}\hat{\mathcal{M}}_{\sigma_i} (1-\hat{\mathcal{M}}_{\sigma_j}).
\end{align}
All together we have
\begin{align}
    Pr[\hat{k}>C] = \sum_{l=C+1}^{N}\sum_{\sigma \in \Sigma_l(N)}\prod_{i=1}^l \prod_{j=l+1}^{N}\hat{\mathcal{M}}_{\sigma_i} (1-\hat{\mathcal{M}}_{\sigma_j}).
\end{align}
Going back to the bound (\ref{app:noisy_sr_bound})
\begin{align}
    \sum_{l=C+1}^{N_{CA}}\sum_{\sigma \in \Sigma_l(N)}\prod_{i=1}^l \prod_{j=l+1}^{N_{CA}}\hat{\mathcal{M}}_{\sigma_i} (1-\hat{\mathcal{M}}_{\sigma_j}) \geq P_{CA}.
\end{align}
Now we assume that all $\hat{\mathcal{M}}_i$'s are independent and identically distributed and introduce the notation \\$\forall_{i=1,\dots,N} \mathbb{E}[\hat{\mathcal{M}}_i] = \mathbb{E}[\hat{\mathcal{M}}_f]$. The separate Bernoulli trials are indeed independent. For them being identically distributed we have to perform a physical assumption that the quantum device does not decalibrate during measurements.
With the above assumption we take take an expectation value with respect to $\hat{\mathcal{M}}_i$'s of both sides of the inequality and obtain
\begin{align}
     P_{CA} \leq \sum_{l=C+1}^{N}\sum_{\sigma \in \Sigma_l(N)}\prod_{i=1}^l \prod_{j=l+1}^{N}\mathbb{E}\left[\hat{\mathcal{M}}_{\sigma_i}\right] (1-\mathbb{E}\left[\hat{\mathcal{M}}_{\sigma_j}\right]) = \sum_{l=C+1}^{N} {N \choose l} \mathbb{E}[\hat{\mathcal{M}}_f]^l(1-\mathbb{E}[\hat{\mathcal{M}}_f])^{N-l}
\end{align}
The function on the right-hand side of the above equality can be identified as a cumulative mass function for a binomial distribution with respect to expectation value of the noisy probability $\hat{\mathcal{M}}_f$. The similar analysis can be done for the probability of obtaining $\hat{k}<C$. Therefore, we come to the conclusion that the bounds (\ref{eq:sr_bound}) for noisy devices have exactly the same form but instead of the true value of probability of success, one has to use the expectation value of the random variable of probability of success.
For example the bound (\ref{eq:sr_fidelity_bound}) for noisy devices has the following form
\begin{align}
    N \geq log_{1-\mathbb{E}[\hat{\kappa}_f]}(1-P_{CA}).
\end{align}

For more concrete analysis of the concentration avoidance bounds one has to assume noise properties of the device.
Assuming the noise channel as depolarizing, and that $\rho_f = (1-p)\rho + p \tilde{\rho}$ we have
\begin{equation}
    \mathbb{E}\left[ \hat{\kappa}_f \right] = \Tr[| 0 \rangle \langle 0 |\rho_f] = (1-p)\kappa + 2^{-n}p
\end{equation}
The $2^{-n}$ scaling is the one corresponding to the exactly Haar-random distributed data points and is expected to be, on average, the strongest scaling for independent kernel values. Therefore, we conclude that considering the noise in the quantum circuits, on average, decreases the resultant, effective kernel values. This increases the number of measurements needed for obtaining reliable results.

\section{Error-Correction Estimates}\label{App:error_estimates}
\subsection{Physical Qubits Count, and Algorithm Runtime}
\label{app:phys_qb_count}

In our work, we considered an implementation of the circuit with quantum error-correction based on the surface code \cite{Fowler2012Sep,Terhal2015Apr}. For this, we used the Microsoft \texttt{Azure Quantum Resource Estimator} \cite{beverland2022assessing} to estimate the total number of qubits required when error-correction is performed (the total number of physical qubits), and the total runtime of the algorithm. The main parameter we have to fix in the resource estimator is the maximum error budget allowed for the circuit. We take it equal to the upper bounds on $p$ from \eqref{eq:final_cond_1} and \eqref{eq:p_projected}. Apart from this parameter, we used the default parameters for the resource estimator, applied on superconducting qubits (qubit name ``qubit\_gate\_ns\_e3", QEC scheme ``surface\_code" in the software parameters). To be more specific, we considered that the physical single and two-qubit Clifford gates, the non-Clifford $T$-gate and the measurements have a probability of error $10^{-3}$. All the physical quantum gates have a duration of $50$ns and a physical measurement lasts for $100$ns. The error-rate of a logical gate takes the expression $p_L=0.03(p_{\text{phys}}/0.01)^{(d+1)/2}$ \cite{beverland2022assessing}, where $p_{\text{phys}}=10^{-3}$ is the probability of error of the physical gates, and $d$ is the code distance. We then estimated the physical resources required to implement the circuits of section \ref{sec:results}. The exact circuits depend on the real parameters chosen of the encoding function: it could also make the physical resource estimate including error-correction depend on this choice. In practice, we numerically checked that, in contrary to the number of shots, the physical resource estimate is actually independent on the exact parameters chosen. We performed multiple physical qubits' and runtime's estimates of quantum state tomography with \texttt{Azure Quantum Resource Estimator} with random values of the ZZ-feature map parameters, for a full spectrum of the studied number of logical qubits. In no case did the result depend on the feature map parameters. In practice, we make use of these resource estimate in Figure \ref{fig:QRE}.

\subsection{Energy Requirements}
The total number of physical qubits (i.e. the actual number of qubits required, including the ones used for error-correction) and the runtime of the algorithm (including the repetition for statistics) allows us to make a simple energy estimate for the algorithm, based on the results of \cite{Fellous-Asiani2023Oct}. Our goal here is to give an indication for the order of magnitude energy required by the quantum computer, rather than a detailed estimate which goes beyond the scope of our paper. In \cite{Fellous-Asiani2023Oct}, a holistic approach, dubbed ``MNR" was proposed to minimize the energy consumption of a quantum computer. It relies on establishing the precise connection between the algorithm's success rate and its energy consumption. Doing so allows to minimize the total energy bill of the computation by modifying all the parameters that can affect the noise: for instance, depending on the algorithm's size, the qubit's temperature will adapt in coordination with the amount of error-correction performed (hence the total number of qubits in the computer) to minimize the consumption while guaranteeing the targetted error-rate for the algorithm. It can sometimes lead to non-trivial effect such as putting the qubits at a hotter temperature, making them noisier, but compensating by doing more error-correction, leading to a computer that is more energy-efficient overall \cite{Fellous-Asiani2023Oct}. The work of \cite{Fellous-Asiani2023Oct} also showed that for a quantum computer based on microwave driven superconducting qubits, for a cryogenics working at Carnot efficiency (this efficiency gives the correct order of magnitude in cryogenics efficiency for a well-optimized large-scale cryogenic unit \cite{Fellous-Asiani2023Oct,Parma2014,martin2022energy}), the power consumption is dominated by the room temperature electronics given state of the art values in CMOS electronics energy consumption (see \cite{Fellous-Asiani2023Oct} to understand all the details behind the model). This consumption is composed of electronics (i) generating and digitizing the microwave signals driving the qubits, or reading their state, (ii) performing the classical computations required by the quantum computer. In \cite{Fellous-Asiani2023Oct}, (ii) had a negligible cost compared to (i) because of the task performed, and the error-correction scheme chosen. Here, we implement another algorithm, and consider another error-correction scheme than the one used in \cite{Fellous-Asiani2023Oct}: we use the surface code \cite{Fowler2012Sep,Terhal2015Apr}. For these reasons, we need to check if neglecting (ii) in the bill is still reasonable. The computations required for (ii) contain: (ii.a) doing the classical computations required by error-correction, (ii.b) doing the classical computations required by the algorithm. To perform the tasks of (i), state of the art CMOS electronics allows for a power budget less than $30$ mW per physical qubit \cite{Park2021Feb,frank2022cryo,Kang2022Aug}. We now argue that all costs for (ii) are negligible compared to this $30$mW per physical qubit. For (ii.a), recent developments indicate classical computations for error-correction having a negligible consumption compared to $30$ mW. Indeed, a quantum memory for a distance $23$ surface code (a larger amount of error-correction than what we need in this paper\footnote{The required code distance range for all investigated problem sizes ranges between $5$ and $17$}) requires about $8 ~ \mu$W per physical qubit \cite{barber2023real}. We are not implementing a quantum memory in this paper, but the total classical computing cost would be similar for the implementation of a non-trivial algorithm based on lattice surgery to implement the logical operations \cite{Kenton_Barnes} (see \cite{horsman2012surface,litinski2019game} for references on lattice surgery, which is the scheme used by the \texttt{Azure Quantum Resource Estimator} \cite{beverland2022assessing}). There is a caveat here: this estimate of $8 ~ \mu$W per physical qubit is based on a decoding algorithm (i.e. the classical algorithm allowing to know which correction should be applied on the qubits when errors are detected) having the same accuracy as the Union Find decoder \cite{delfosse2021almost,Kenton_Barnes}. Our estimates in qubit's number are using the \texttt{Azure Quantum Resource Estimator} which implicitly assumes another decoding algorithm \cite{beverland2022assessing} (changing this algorithm can change the logical error rates, hence the number of qubits required for error-correction). Hence, we assume in our estimates that the consumption of this decoding algorithm will not be $\approx 1000 \times$ bigger than the one of the Union Find decoder so that this cost is still negligible. We estimate this assumption to be reasonable, but we leave for future work a detailed analysis of this question. For (ii.b), the classical computation we need to do consists of rearranging the estimated reduced density matrices into the Gram matrix. Although the number of kernel matrix entries scale quadratically with the size of the data set, the single kernel matrix entry computation requires a small amount of simple algebraic operations (about $10n$ computations of sum, square or factor of a floating point number and one exponentiation per kernel entry) and therefore can be neglected in our considerations. All this explains why $30$ mW of consumption per physical qubit is an appropriate estimate for the quantum computer's consumption.  







\section{Datasets}\label{App:datasets}

\begin{table}[h!]
\caption{The summary of the datasets used. The $\#$ of subsets indicated whether the dataset was stratified and into how many subsets, $m$ is the number of data points in the dataset/each subset, while $m_0$ and $m_1$ indicate how many patterns belong to each of two classes, we also report the imbalance ratio (IR), and the number of features ($n$).}
\label{tab:datasets}
\centering
\begin{tabular}{l|lllllll}
dataset      & ref.             & $\#$ of subsets & $m$    & $m_0$ & $m_1$ & IR & $n$   \\ \hline
Sonar        & \cite{sonar}    & 1               & $208$  & $111$ & $97$  & $1.14$          & $60$  \\
Twonorm      & \cite{twonorm}  & 74              & $100$  & $50$  & $50$  & $1.00$          & $20$  \\
Indian Pines & \cite{PURR1947} & 1               & $1460$ & $730$ & $730$ & $1.00$          & $200$
\end{tabular}
\end{table}

\noindent For the numerical study we have chosen three datasets, which are summarized in the Table \ref{tab:datasets}. Additionally, below we provide a short description of each dataset, with pre-processing discussed. The prepared datasets are also provided in the project's repository \cite{github}.

Here, we also point out that the studied datasets do not impose training-test splits. In this work we are mainly interested in the properties of the kernel matrices for the dataset, rather than the machine learning performance of the trained models. Therefore, we do not introduce any additional splits into the training and test sets, we obtain kernel matrices for the whole datasets, or for the stratified datasets.

\subsection{Connectionist Bench (Sonar, Mines vs. Rocks)}
Following \cite{sonar}, the task in the dataset is to train a model to discriminate between sonar signals bounced off a metal cylinder and those bounced off a roughly cylindrical rock.
It consists of the $208$ data points, divided into two classes of $111$ and $97$ patterns, each with $60$ features. 
The applied pre-processing consists of centering and normalization of the data. The subsequent features for the $n$-dependent analysis are picked in the order of descending variance.

\noindent This is the main dataset of our work and is used for visualization of the issues studied.
It fits the study particularly well, as the size of the dataset does not prohibit us from performing extensive simulations for multiple kernel families and feature map hyperparameters. It is also an almost balanced data set (imbalance ratio of $1.14$) which makes it suitable for the analysis using median as a measure of statistical behavior of kernel matrix entries.

\subsection{Twonorm}
The artificial dataset \cite{twonorm} consisting of $7400$ patterns. The data points are distributed into two, equal classes with $3700$ points each and with $20$ features. The dataset is too big to perform quantum simulations on it. Therefore, the dataset was stratified into $74$ balanced subsets, each consisting of $100$-points, the kernel simulation is performed every subset separately. Together, the subsets cover the whole dataset.
The applied pre-processing consists of centering and normalization of the data.

\subsection{Indian Pines}
The real-life dataset consisting of a scene AVIRIS sensor \cite{PURR1947} over the Indian Pines test site in North-western Indiana. The scene is taken with the sensor in range of the wavelengths spanning $0.4–2.5 10^{-6}m$, which are then gathered into $200$ features. It covers agriculture areas and serves as a dataset for multiclass classification problems. In our case, we take only two classes of pixels: \textit{Corn-mintill} and \textit{Grass-trees}. We artificially reduce the dataset while balancing the classes by removing $100$ points from the \textit{Corn-mintill} class. We end up with total number of $1460$ data points, each class consisting of $730$ points.
Again, the applied pre-processing consists of centering and normalization of the data.


\end{document}